\documentclass[a4paper,11pt]{article}
\usepackage{jcappub} 
\usepackage{lineno}
\usepackage{comment}

\arxivnumber{1234.56789} 
\title{\boldmath Standard Perturbation Theory for Interacting Dark Sector cosmologies I: Breakdown of Einstein–de Sitter kernels}

\author[a]{Matheus Wolney,}
\author[a]{\ Gabriel Sampaio,}
\author[a]{\ Humberto Borges,}
\author[b]{\ Iuri Baranov,}
\author[a,c,1]{\ Rodrigo von Marttens\note{Corresponding author.}}

\affiliation[a]{Instituto de Física, Universidade Federal da Bahia, Salvador - BA, 40210-340, Brasil}
\affiliation[b]{Instituto Federal de Educação, Ciência e Tecnologia da Bahia, 40301-015, Salvador, BA, Brazil}
\affiliation[c]{PPGCosmo, Universidade Federal do Espírito Santo, Vitória-ES, 29075-910, Brazil}

\emailAdd{rodrigovonmarttens@gmail.com}

\abstract{Interacting dark sector (IDS) models provide a commonly explored extension of the standard $\Lambda$CDM cosmology, allowing for non-gravitational energy--momentum exchange between cold dark matter (CDM) and dark energy (DE). Although such models can be constructed to reproduce the same background expansion history as $\Lambda$CDM, their impact on the growth of cosmic structures is fundamentally different and requires a careful treatment of cosmological perturbations. In this work, we develop the one-loop Standard Perturbation Theory (SPT) formalism for IDS cosmologies without invoking the Einstein--de~Sitter (EdS) approximation. We show that even weak dark sector interactions induce a non-trivial time dependence in the perturbative kernels, leading to a breakdown of the EdS approximation commonly assumed in $\Lambda$CDM analyses. By deriving and numerically solving the evolution equations for the second- and third-order kernels, we compute the corresponding one-loop corrections to the matter power spectrum and find that the resulting deviations can significantly exceed the percent level, even for small interaction strengths. Our results demonstrate that nonlinear corrections are systematically enhanced in IDS models and that neglecting the full time dependence of the kernels can lead to biased predictions on mildly nonlinear scales. These findings establish the necessity of a time-dependent perturbative treatment for IDS scenarios and provide a robust framework for precision tests using nonlinear large-scale structure (LSS) observables.}

\begin{document}
\maketitle
\flushbottom

\section{Introduction}
\label{sec:intro}

The $\Lambda$CDM model has established itself as the standard framework of modern cosmology, providing an excellent fit for a wide range of observations, from the cosmic microwave background (CMB) to large-scale structure (LSS)~\citep{Heavens:2008zz}. However, the increasing precision of cosmological data has exposed a growing number of internal tensions that challenge the completeness of this model~\citep{Perivolaropoulos:2021jda}. The most notable example is the $H_0$ tension, i.e., the persistent discrepancy between early- and late-Universe determinations of the Hubble constant~\citep{Bernal:2016gxb}. More recently, still within the $\Lambda$CDM framework, baryon acoustic oscillation (BAO) measurements from DESI~\citep{DESI:2025zgx} and Type~Ia supernova observations~\citep{Scolnic:2021amr,Rubin:2023jdq,DES:2024jxu} have delivered conflicting constraints on the matter density parameter $\Omega_{\rm m}$, revealing a growing tension between late-time probes. In addition, BAO analyses from DESI have shown a preference for departures from a cosmological constant, pointing toward the possible presence of dynamical dark energy. Taken together, these results motivate the exploration of extensions of the $\Lambda$CDM model that go beyond a constant vacuum energy component.

In this context, several departures from the $\Lambda$CDM paradigm can be considered~\citep{DiValentino:2021izs}. Among the most widely studied possibilities are models with dynamical dark energy~\citep{deSouza:2025rhv}, theories of modified gravity~\citep{Chudaykin:2024gol}, and scenarios in which dark matter is non-minimally coupled to dark energy~\citep{Petri:2025swg}. In this work, we focus on the latter class of models, characterized by a non-gravitational interaction between dark energy (DE) and cold dark matter (CDM). Indeed, a fundamental obstacle in identifying the physical nature of the dark sector is the so-called \emph{dark degeneracy}~\cite{vonMarttens:2019ixw}. For instance, at the level of a homogeneous and isotropic background, it is well known that distance-based observations alone cannot distinguish between dynamical dark energy models and scenarios in which dark matter and dark energy interact through non-gravitational couplings~\cite{vonMarttens:2022xyr}. In fact, for any background evolution generated by a given dynamical dark energy equation of state, one can always build an interacting dark sector (IDS) model that reproduces the same expansion history~\cite{vonMarttens:2019ixw}. This degeneracy has also been explored in the context of non-minimal couplings within the dark sector using model-independent approaches~\citep{vonMarttens:2018bvz,vonMarttens:2020apn}.

At the level of perturbations, this degeneracy may persist or be broken depending on the physical properties of the dark energy component, in particular on the choice of its sound speed. While convenient choices of the dark energy sound speed can preserve the degeneracy, it is well known that for a canonical scalar field with a luminal sound speed the degeneracy is generically broken~\cite{vonMarttens:2019ixw}. This observation highlights that background observables alone are insufficient to unveil the nature of the dark sector and that cosmological perturbations provide a crucial and promising avenue to probe the underlying physical mechanisms driving cosmic acceleration.

Within this perspective, breaking the dark degeneracy necessarily requires going beyond background cosmology and exploiting the wealth of information encoded in cosmological perturbations. Current and forthcoming surveys deliver high-precision measurements of the growth of cosmic structures, redshift-space distortions, weak gravitational lensing, and higher-order clustering statistics~\citep{EUCLID:2011zbd,LSST:2008ijt}. These observables are directly sensitive to the dynamics of perturbations in the dark sector and, in particular, to possible non-standard Physics. An accurate theoretical description of structure formation in the mildly non-linear regime is therefore essential. This regime is especially relevant for large-scale structure analyses, as it contains a substantial fraction of the cosmological information accessible to perturbative approaches and plays a central role in discriminating the nature of the dark components.

Within the framework of cosmological perturbation theory, Standard Perturbation Theory (SPT) provides a well-established analytical tool for describing the evolution of density and velocity perturbations beyond the linear regime~\cite{Bernardeau:2001qr}. In the context of $\Lambda$CDM and other minimally coupled cosmologies, it is common practice to adopt the Einstein--de Sitter (EdS) approximation, in which the time dependence of higher-order perturbations factorizes and the perturbative kernels depend solely on the configuration of wavevectors. Remarkably, this approximation is known to remain highly accurate for $\Lambda$CDM, despite the presence of a cosmological constant~\citep{Pietroni:2008jx}.

This property, however, is not generic. 
When a non-gravitational coupling between dark matter and dark energy is introduced, the individual conservation equations are modified, giving rise to additional terms in the continuity and Euler equations that characterize the energy and momentum transfer within the dark sector~\cite{Benetti:2019lxu}. These contributions introduce a non-trivial time dependence and alter the coupling between density and velocity perturbations, thereby violating the assumptions underlying the EdS approximation. As a consequence, the use of EdS kernels in IDS cosmologies is not formally justified and can lead to biased predictions for non-linear observables. Although IDS models typically recover the $\Lambda$CDM limit at sufficiently high redshifts, where the interaction becomes dynamically negligible, the main issue arises at low and intermediate redshifts, where the dark sector interaction becomes increasingly relevant and significantly alters the time evolution of perturbations~\cite{vonMarttens:2018iav}.

In this work, we develop the one-loop SPT framework for IDS models without imposing the EdS approximation. We derive the evolution equations for the perturbative kernels directly from the modified continuity and Euler equations, explicitly accounting for their non-trivial time dependence induced by dark sector interactions. Using these kernels, we compute the corresponding non-linear matter power spectrum and assess the impact of the interaction on mildly non-linear scales.

Although the EdS approximation might provide a first approximation for IDS models, the level of precision achieved by current observations demands a more careful assessment. Recent results from DESI and other LSS surveys require alternative cosmological models to be tested with an accuracy comparable to that achieved for $\Lambda$CDM. This is especially important in view of the fact that perturbative corrections in the mildly non-linear regime can play a decisive role in discriminating between competing dark sector scenarios. Our analysis therefore aims to quantify the validity of the EdS approximation in interacting dark sector cosmologies and to establish whether its use is compatible with the precision required by present and forthcoming observational data.

The work is organized as follows. In Sec.~\ref{sec:ids_models}, we introduce the IDS framework, presenting the background evolution equations and deriving the modified Boltzmann–Poisson system. In Sec.~\ref{sec:ids_spt}, we develop the SPT formalism for IDS cosmologies, explicitly deriving the evolution equations for the perturbative kernels up to third order and generalizing the construction to arbitrary order. In Sec.~\ref{sec:results}, we present our numerical results, first analyzing the time dependence of the second- and third-order kernels and then quantifying the impact of this time dependence on the one-loop matter power spectrum. Finally, in Sec.~\ref{sec:conclusions}, we summarize our main conclusions and outline directions for future work. Additional triangle configurations for the kernels are discussed in Appendix \ref{ap:kernel}.

\section{Interacting Dark Sector models}
\label{sec:ids_models}

A fundamental assumption within the standard cosmological model posits that, at late-times, all constituents of the Universe undergo time-evolutions independently of one another. This means that their late-time evolution is solely affected by gravity, and they do not interact with each other \cite{amendola2010dark}. From a formal perspective, this hypothesis manifests itself in the condition that the covariant derivative of individual energy-momentum tensor is zero for each component independently. We keep this hypothesis for baryons\footnote{Since we are interested only in the late-time evolution of the Universe, from now on we neglect the radiation component.}, while relaxing it for the constituents within the dark sector of the Universe, considering that they are capable of interacting with one another. Regarding the baryonic component, denoted by the lower index ${\rm b}$, its time evolution is obtained through the usual energy conservation,
\begin{equation} \label{eq:db} 
    \dot{\rho}_{\rm b}+3H\rho_{\rm b}=0 \,. 
\end{equation}
On the other hand, in order to formulate mathematically the dark sector interaction, we consider that the covariant derivatives of CDM and DE are not zero, instead, they are related to a source four-vector \cite{valiviita2008large},
\begin{equation} \label{eq:dTmunu}
    \nabla_{\mu}T^{\mu\nu}_{\Lambda}=-\nabla_{\mu}T^{\mu\nu}_{\rm c}=Q^{\nu} \,,
\end{equation}
where the subscripts $\Lambda$ and ${\rm c}$ denote the dark energy (DE) and cold dark matter (CDM) components, respectively\footnote{We adopt the notation $\Lambda$ for the dark energy component since we use an equation of state $w=-1$.}.

Regarding Eq.~\eqref{eq:dTmunu}, there are two noteworthy aspects to address. Firstly, within the framework of General Relativity, Bianchi identities ensures that the Universe as a whole must be a conservative system, meaning that the covariant derivative of the total energy-momentum tensor, formed by the sum of individual component tensors, must equate to zero \cite{mukhanov2005physical}. In order to satisfy this conservation requirement, the introduced source term must influence the dynamics of interacting components with opposite effects, ensuring that their cumulative contribution remains zero. Second, given our current lack of comprehensive understanding regarding fundamental properties of CDM and DE, deriving an explicit form for the source term based on first principles remains implausible. The usual approach involves adopting phenomenological analytical expressions for the source term, which characterizes a specific IDS model \citep{clemson2012interacting, valiviita2010observational}.

At the background level, the time component of Eq.~\eqref{eq:dTmunu} establishes the following equations for the energy balance of the dark sector components,
\begin{eqnarray} 
    \dot{\rho}_{\rm c}+3H\rho_{\rm c}&=&Q \,, \label{eq:dcdm} \\
    \dot{\rho}_{\Lambda}&=&-Q\,. \label{eq:dde}
\end{eqnarray}
where the dot denotes the partial derivative with respect to time and $Q$ denotes the background time component of the source four-vector. The Eqs.~\eqref{eq:dcdm} and~\eqref{eq:dde} serve as formal description of the energy exchange occurring in the dark sector as a consequence of the interaction. The sign of the source term defines the direction of energy transfer: when $Q$ is positive, energy is transferred from DE to CDM, whereas a negative $Q$ implies energy transfer in the opposite direction. Also, considering the background level, the characterization of the interacting components as perfect fluids implies an identically zero transfer of momentum. Several models have been proposed in the literature, alongside the application of different observational tests to constrain how dynamically relevant these interactions can be \citep{figueruelo2026late,li2024constraints,borges2023testing}. Furthermore, it is important to mention that IDS models have been proposed as viable alternatives to address the cosmological tensions~\citep{DiValentino:2020zio,DiValentino:2020vvd}.

In the subsequent sections, our development will primarily focus on the late-time perturbations of the total matter, which comprises both CDM and baryonic matter. In the context of the $\Lambda$CDM model, this is not an issue since at low redshifts the baryons are already ``captured'' by the CDM potential wells, so that their fluctuations evolve together. Nevertheless, in our specific scenario, the relation between the fluctuations of total matter, CDM, and baryonic matter is not trivial. The complexity here arises due to the new dynamics introduced by the interaction in the dark sector, which does not affect baryons~\cite{vomMarttens:2014ftc}. Even though the forthcoming development emphasizes the total matter perturbations, in certain relevant situations, we will also examine some cases in which the separation of these components becomes pertinent. Given the absence of pressure in both CDM and baryons, the decomposition of total matter, denoted by the lower index $m$, can be easily done using the following relations,
\begin{eqnarray}
    \rho_{\rm m}&=&\rho_{\rm c}+\rho_{\rm b} \,, \label{eq:rhom} \\
    \rho_{\rm m} u_{\rm m}^{i}&=&\rho_{\rm c} u_{\rm c}^{i}+\rho_{\rm b} u_{\rm b}^{i} \,. \label{eq:rhoum}
\end{eqnarray}

\subsection{The Boltzmann-Poisson system}
\label{ssec:Boltzmann-Poisson}

Starting from the Boltzmann equation and taking the small-scale, non-relativistic limit appropriate for structure formation, one obtains the Boltzmann equation for the evolution of the matter phase-space distribution, 
\begin{equation} \label{eq:vlasov}
    \frac{\partial f_{\rm m}}{\partial t}+\frac{\partial f_{\rm m}}{\partial x_{\rm m}^{i}}\frac{p_{\rm m}^{i}}{am}-\frac{\partial f_{\rm m}}{\partial p_{\rm m}^{i}}\left(Hp_{\rm m}^{i}+\frac{m}{a}\partial^{i}\Psi\right)=C\left[f_{\rm m}\right]\,.
\end{equation}

Compared to the $\Lambda$CDM case, the Boltzmann equation now contains an additional source term, denoted by, $C\left[f_{\rm m}\right]$,  arising from the non-gravitational interaction in the dark sector. As will be discussed below, this extra contribution can be consistently related to the interaction source $Q^{\mu}$, which describes the exchange of energy and momentum between dark matter and dark energy at the level of the energy--momentum tensor. 

\subsubsection{Energy-momentum balance}
\label{sssec:energy-momentum}

The energy conservation equation follows from the zeroth moment of the Boltzmann equation, which is obtained by taking the momentum average of all terms in Eq.~\eqref{eq:vlasov}. Accordingly, multiplying Eq.~\eqref{eq:vlasov} by the particle energy and taking the zeroth moment leads to the energy balance equation for the total matter component,
\begin{equation} \label{eq:mzero}
    \frac{\partial \rho_{\rm m}}{\partial t}+\frac{1}{a}\partial_{i}\left(\rho_{\rm m} u_{\rm m}^{i}\right)+3H\rho_{\rm m}=\int\frac{d^{3}p_{\rm m}}{\left(2\pi\right)^{3}}C\left[f_{\rm m}\right]E_{\rm m} \,.
\end{equation}
Note that, for the matter component, the slow-motion approximation has been adopted, such that $E_{\rm m} \approx m$. The first two terms on the left-hand side of Eq.~\eqref{eq:mzero} represent the local time evolution and the flux of the matter density, respectively, while the third term describes the dilution of the density due to cosmological expansion.

We now turn to the first moment of the Boltzmann equation, obtained by multiplying Eq.~\eqref{eq:vlasov} by the three-momentum $p_{\rm m}^{i}$ and taking the momentum average. This yields
\begin{equation} \label{eq:mone}
    \frac{\partial\left(\rho_{\rm m} u_{\rm m}^{i}\right)}{\partial t}+\frac{1}{a}\partial_{j}\left(\rho_{\rm m}u_{\rm m}^{j}\right)u_{\rm m}^{i}+\frac{1}{a}\rho_{\rm m}u_{\rm m}^{j}\partial_{j}u_{\rm m}^{i}+\frac{1}{a}\partial_{j}\sigma_{\rm m}^{ij}+4H\rho_{\rm m}u_{\rm m}^{i}+\frac{1}{a}\rho_{\rm m}\partial^{i}\Psi=\int\frac{d^{3}p_{\rm m}}{\left(2\pi\right)^{3}}\ C\left[f_{\rm m}\right]p_{\rm m}^{i}
\end{equation}
where $\sigma_{\rm m}^{ij}$ denotes the stress tensor of the matter component, defined in terms of the second moment of the distribution function as
\begin{equation} \label{eq:sigmaij}
    \sigma_{\rm m}^{ij}\equiv\rho_{\rm m}u_{\rm m}^{i}u_{\rm m}^{j}-\frac{1}{m}\int\frac{d^{3}p_{\rm m}}{\left(2\pi\right)^{3}}\ p_{\rm m}^{i}p_{\rm m}^{j} f_{\rm m}\,.
\end{equation}

Concerning the stress tensor, which encodes velocity–dispersion effects, it may in general give rise to isotropic pressure and anisotropic shear stresses. In the CDM case, however, these contributions are negligible and, as in the $\Lambda$CDM model, the matter stress tensor is assumed to vanish. This assumption is also adopted here for IDS models. This means that we consider that the non-gravitational interaction between the dark components is taken to modify the energy–momentum exchange at the fluid level without inducing significant velocity dispersion in the matter component. 

The Euler equation, which governs the momentum balance of the matter fluid, is obtained by combining Eqs.~\eqref{eq:mone} and~\eqref{eq:mzero} so as to eliminate the explicit time derivative of the density. For subsequent perturbative analyses, it is convenient to recast the resulting equation in terms of the velocity divergence, defined as $\theta_{\rm m}\equiv\partial_i u_{\rm m}^{i}$. Taking the spatial divergence of the Euler equation then yields
\begin{equation} \label{eq:euler}
\frac{\partial \theta_{\rm m}}{\partial t} + H \theta_{\rm m} + \frac{1}{a}\partial_i\!\left(u_{\rm m}^{j}\partial_{j}u_{\rm m}^{i}\right) + \frac{1}{a}\nabla^{2}\Psi = \frac{1}{\rho_{\rm m}} \partial_i \left[\int \frac{d^{3}p_{\rm m}}{(2\pi)^{3}}\, C[f_{\rm m}]\, p_{\rm m}^{i} - u_{\rm m}^{i}\int \frac{d^{3}p_{\rm m}}{(2\pi)^{3}}\, C[f_{\rm m}]\, E_{\rm m} \right]\,.
\end{equation}
Each term on the right-hand side of Eq.~\eqref{eq:euler} contributes a distinct component to the acceleration of the fluid. The momentum-weighted collision term represents a direct dynamical force acting on the fluid, whereas the energy-weighted term accounts for the acceleration associated with variations in the inertia of the fluid, resulting from energy exchange.

\subsubsection{The interaction term}
\label{sssec:qterm}

To establish a consistent connection between the kinetic description and the fluid formalism, we project the energy–momentum balance equation for the dark sector along and orthogonal to the matter four-velocity. In the sub-horizon limit and adopting the longitudinal Newtonian gauge, the projection parallel to the four-velocity identifies the energy transfer rate along the fluid worldline, leading to the continuity equation given in Eq.~\eqref{eq:mzero}, allowing us to write the time-component of the interaction source term as,
\begin{equation} \label{eq:defQ}
    \int\frac{d^{3}p_{\rm m}}{(2\pi)^{3}}\, C[f_{\rm m}]\,E_{\rm m}\equiv-u_{\rm m}^{\mu}Q_{\mu}=Q\,.
\end{equation}
Analogously, the projection onto the hypersurface orthogonal to the matter four-velocity yields the local rate of momentum transfer between the dark-sector components. At the kinetic level, this quantity is defined through the momentum-weighted collision integral,
\begin{equation} \label{eq:Qi}
    \int\frac{d^{3}p_{\rm m}}{(2\pi)^{3}}\, C[f_{\rm m}]\,p_{\rm m}^{i}\equiv Q^{i}\,.
\end{equation}
which corresponds to the force density appearing in the Euler equation, as introduced in Eq.~\eqref{eq:euler}. Taken together, the projections parallel and orthogonal to the four-velocity consistently recover the modified Boltzmann–Poisson system. Within this framework, the energy-weighted contribution in Eq.~\eqref{eq:euler} reflects the adjustment to the momentum balance associated with mass exchange along the fluid worldline, whereas the momentum-weighted term represents a direct dynamical force. This identification ensures that the energy transfer rate reproduces the expected phenomenological background behavior of IDS models while remaining consistent with the local conservation laws governing the perturbations.

For the momentum balance, we adopt the geodesic interaction hypothesis, which assumes that the interaction four-vector is aligned with the matter flow, i.e. $Q^{\mu} = Q u^{\mu}_{\rm m}$. Under this assumption, the direct momentum transfer exactly compensates the inertial change associated with energy exchange, causing the right-hand side of Eq.~\eqref{eq:euler} to vanish. Consequently, matter particles continue to follow geodesics and the Euler equation retains its standard form. In addition, we assume that the interaction rate does not develop local perturbations, so that its perturbation can be neglected ($\delta Q \approx 0$ at all orders). Physically, this assumption can be motivated by a spatially uniform collision rate, implying that the energy exchange is governed entirely by the background evolution rather than by local fluctuations.

With the interaction consistently defined in terms of the source four-vector, the energy conservation and Euler equations can be rewritten by expressing the kinetic source terms in terms of the scalar energy-transfer function $Q$ and its associated momentum-transfer components $Q^{i}$:
\begin{eqnarray}
    \frac{\partial \rho_{\rm m}}{\partial t}+\frac{1}{a}\partial_{i}\left(\rho_{\rm m} u_{\rm m}^{i}\right)+3H\rho_{\rm m}&=&Q\,, \label{eq:energyQ}\\
    \frac{\partial \theta_{\rm m}}{\partial t} + H \theta_{\rm m} + \frac{1}{a}\partial_i\!\left(u_{\rm m}^{j}\partial_{j}u_{\rm m}^{i}\right) + \frac{1}{a}\nabla^{2}\Psi &=& 0\,. \label{eq:eulerQ} 
\end{eqnarray}
    Up to this stage, the dark sector interaction has been treated in full generality through an arbitrary four-vector $Q^{\mu}$. 
Although this general formulation is useful for identifying all possible physical effects, Eq.~\eqref{eq:eulerQ} shows explicitly that interaction terms might introduce scale-dependent contributions to the perturbation dynamics already at linear order. To proceed, we adopt a specific interaction model, which will be important to provide quantitative results. The chosen model is characterized by,
\begin{equation} 
\label{eq:Qmodel_bg}
    Q=3H\gamma\rho_{\Lambda}\,,
\end{equation}
where, from now on the function $Q$ denotes only the background contribution of the CDM-DE coupling, whereas perturbations will be denoted by $\delta Q$. This parametrization has been widely explored in the literature and has been discussed as a phenomenologically viable mechanism that may help alleviate current cosmological tensions~\citep{DiValentino:2017iww}.

\subsection{Beyond background level}
\label{ssec:beyondbg}

The study of non-linear structure formation necessarily requires going beyond the homogeneous and isotropic background description of the Universe. While background observables probe the global expansion history, the growth of cosmic structures is governed by the dynamics of perturbations around this background. Standard Perturbation Theory (SPT) provides a systematic framework to describe this evolution on mildly non-linear scales, where perturbative methods remain valid and most of the cosmological information from LSS surveys is encoded.

In this context, matter inhomogeneities are conveniently characterized in terms of the density contrast, defined as the fractional deviation of the local matter energy density from its homogeneous background value, $\delta_{\rm m}(t,\vec{x})\equiv \delta\rho_{\rm m}(t,\vec{x})/\rho_{\rm m}(t)$, which physically quantifies the degree of matter clustering. With this definition, the full matter energy density can be expressed as
\begin{equation} \label{eq:rhomgen}
\rho_{\rm m}(t,\vec{x})\;\longrightarrow\;\rho_{\rm m}(t)\left[1+\delta_{\rm m}(t,\vec{x})\right].
\end{equation}
The density contrast is therefore the fundamental quantity driving the gravitational instability that leads to structure formation. On sub-horizon scales, where relativistic effects are negligible, the gravitational potential generated by these matter inhomogeneities is well described by the Poisson equation,
\begin{equation} \label{eq:poisson}
\nabla^{2}\Psi=-4\pi G a^{2}\rho_{\rm m}\delta_{\rm m}\,,
\end{equation}
where $\nabla^{2}$ denotes the spatial Laplacian. This relation implies that, on small scales, the gravitational potential is suppressed relative to the density contrast by a factor of order $(aH/k)^{2}$. As a consequence, metric perturbations contribute negligibly to the time component of the matter four-velocity at the perturbative level. Subtracting the background contributions from the Eqs.~\eqref{eq:energyQ} and~\eqref{eq:eulerQ}, the evolution of matter perturbations in IDS models can be written as

\begin{eqnarray}
\delta^{\prime}_{\rm m} + \theta_{\rm m} + \mathcal{H} g\, \delta_{\rm m}
&=& - u_{\rm m}^{i}\partial_{i}\delta_{\rm m} - \delta_{\rm m}\theta_{\rm m} \,, \label{eq:continuity_ids} \\
\theta^{\prime}_{\rm m} + \mathcal{H}\theta_{\rm m} + \frac{3}{2}\mathcal{H}^{2}\Omega_{\rm m}\delta_{\rm m}
&=& -\left(\partial_{i}u_{\rm m}^{j}\right)\left(\partial_{j}u_{\rm m}^{i}\right)
- u_{\rm m}^{i}\partial_{i}\theta_{\rm m} \,. \label{eq:euler_ids}
\end{eqnarray}
Here, a prime denotes differentiation with respect to conformal time, and $\mathcal{H}\equiv a'/a$ is the Hubble expansion rate in conformal time. For convenience, we have introduced the dimensionless interaction function \cite{Borges:2017jvi}
\begin{equation} \label{eq:gdef}
g \equiv \frac{aQ}{\mathcal{H}\rho_{\rm m}}\,,
\end{equation}
which parametrizes, at the background level, the strength of the dark sector interaction relative to the Hubble expansion rate and the background matter density. 
With this definition, the interaction enters Eq.~\eqref{eq:continuity_ids} through an additional background-dependent term, which has no analogue in the $\Lambda$CDM model. 
This extra contribution modifies the effective dilution rate of matter perturbations as a direct consequence of the energy exchange between dark matter and dark energy, altering the evolution of the matter density contrast even in the linear regime.

In contrast, within the geodesic interaction scenario considered here, the Euler equation~\eqref{eq:euler_ids} retains the same structure as in $\Lambda$CDM, reflecting the absence of a direct momentum transfer to the matter component. The right-hand sides of Eqs.~\eqref{eq:continuity_ids} and~\eqref{eq:euler_ids} contain the non-linear mode-coupling terms familiar from SPT, which describe the advection of density and velocity perturbations by the matter flow and are responsible for the transfer of power between Fourier modes in the mildly non-linear regime.

Within the framework of SPT, all dynamical quantities are expanded order by order in the amplitude of the fluctuations, under the assumption that perturbations remain sufficiently small on the mildly non-linear scales of interest. This procedure allows the evolution equations to be solved iteratively, with higher-order contributions capturing non-linear mode coupling. In particular, we expand the matter density contrast, velocity divergence, and velocity field as
\begin{equation}
\left\{
\begin{array}{rcl}
\delta_{\rm m} &=& \delta_{\rm m}^{(1)}+\delta_{\rm m}^{(2)}+\delta_{\rm m}^{(3)}+\ldots \vspace{2mm} \\
\theta_{\rm m} &=& \theta_{\rm m}^{(1)}+\theta_{\rm m}^{(2)}+\theta_{\rm m}^{(3)}+\ldots \vspace{2mm} \\
u_{\rm m}^{i} &=& u_{\rm m}^{i(1)}+u_{\rm m}^{i(2)}+u_{\rm m}^{i(3)}+\ldots
\end{array}
\right.
\end{equation}
where the superscript $(n)$ denotes the $n$-th order perturbative contribution. At each order, the evolution of these fields is sourced by mode-coupling terms involving lower-order perturbations. In interacting dark sector models, the interaction term itself may also acquire perturbations, as
\begin{equation} \label{eq:Q_pert}
Q = Q + \delta Q^{(1)} + \delta Q^{(2)} + \delta Q^{(3)} + \ldots\,.
\end{equation}

Although, in principle, perturbations of the interaction term can be consistently incorporated into the analysis, as indicated in Eq.~\eqref{eq:Q_pert}, in this work we adopt the simplifying assumption that dark energy does not cluster on sub-horizon scales. Within the context of our specific interaction model introduced in Eq.~\eqref{eq:Qmodel_bg}, this implies that fluctuations in the interaction term are also strongly suppressed, leading to $\delta Q \approx 0$ at all perturbative orders. This assumption is physically motivated because we consider a luminal effective sound speed, which efficiently suppresses the growth of inhomogeneities at late times. As a result, dark energy remains nearly homogeneous and mainly affects cosmic evolution through the background expansion, rather than directly participating in structure formation.

The interaction operates predominantly at the background level, modifying the global energy exchange within the dark sector while avoiding additional clustering effects in the perturbative dynamics of matter. This perturbative framework therefore provides the starting point for deriving the evolution equations of the standard perturbation theory (SPT) kernels in interacting dark sector models and for assessing how the presence of an interaction alters their time dependence and nonlinear contributions.

Throughout this work, all analyses and figures are performed using a fixed set of cosmological parameters. For the six standard $\Lambda$CDM parameters, we adopt the best-fit values from the Planck 2018 analysis~\citep{Planck:2018vyg}, which define our reference cosmology. For the interacting dark sector (IDS) scenario, we employ the parametrization introduced in Eq.~\eqref{eq:Qmodel_bg} and fix the interaction strength to $\gamma=0.05$. This value lies well within current observational bounds and remains fully consistent with recent statistical analyses based on the latest cosmological datasets~\citep{kaeonikhom2023observational}. Rather than performing an extensive parameter exploration, our objective is to illustrate that even a ``weak'' dark sector coupling can produce physically meaningful modifications in the nonlinear evolution of matter perturbations.

\section{The Standard Perturbation Theory for IDS models}
\label{sec:ids_spt}

The Standard Perturbation Theory (SPT) provides a well-established and widely used framework for describing the non-linear evolution of cosmological density perturbations and has been extensively discussed in the literature~\citep{Bernardeau:2001qr}. In most applications, the formalism is introduced directly at arbitrary perturbative order and formulated in Fourier space, where translational invariance allows the dynamics to be expressed in terms of compact recursive relations for the perturbative kernels. This approach is particularly efficient in the standard $\Lambda$CDM scenario, where the Einstein--de Sitter--like assumptions are suitable.

On the other hand, for IDS cosmologies, the presence of non-gravitational interactions between CDM and DE modifies the continuity and Euler equations already at the linear level and introduces additional time-dependent contributions whose physical origin and dynamical impact are not immediately transparent in a fully general, all-order formulation. For this reason, it is instructive to explicitly examine the perturbation equations at low orders, where the role of the interaction terms can be clearly identified and their propagation to higher orders can be traced unambiguously.

In this section, we therefore derive the perturbation equations explicitly up to third order in Fourier space. This is the minimal order required to compute the matter power spectrum consistently at one-loop level and allows us to isolate the interaction-induced modifications to the standard SPT kernels and their time dependence. This step-by-step construction provides a clear bridge between the linear theory and the fully nonlinear regime and sets the stage for the one-loop analysis presented in the following sections.

\subsection{First-order perturbations}
\label{ssec:first}

As a general rule in cosmological perturbation theory, it is convenient to work in Fourier space, where statistical homogeneity and isotropy imply that first-order perturbations with different wavenumbers evolve independently at linear order. This representation simplifies the analysis and makes the formalism more natural for higher-order perturbative calculations, where mode coupling becomes essential. At first order, the Eqs.~\eqref{eq:continuity_ids} and~\eqref{eq:euler_ids} in Fourier space are
\begin{eqnarray}
\delta^{(1)\prime}_{\vec k}+\theta^{(1)}_{\vec k} + \mathcal{H}g\,\delta^{(1)}_{\vec k} &=& 0 \,, \label{eq:delta1_linear_k}\\
\theta^{(1)\prime}_{\vec k}+\mathcal{H}\theta^{(1)}_{\vec k} + \frac{3}{2}\mathcal{H}^2\Omega_{\rm m}\,\delta^{(1)}_{\vec k} &=& 0 \,, \label{eq:theta1_linear_k}
\end{eqnarray}
where $\delta^{(1)}_{\vec k}$ and $\theta^{(1)}_{\vec k}$ denote the Fourier components of the matter density contrast and velocity divergence, respectively.

Since each Fourier mode evolves independently, the dynamics reduces to a system of ordinary differential equations in time for each wavenumber. A standard approach to study the linear evolution is therefore to combine Eqs.~\eqref{eq:delta1_linear_k} and~\eqref{eq:theta1_linear_k} into a single second-order differential equation for the density contrast~\cite{Velten:2015qua}. In the IDS case, this procedure leads to
\begin{equation} \label{eq:ddelta_ids_linear_k}
\delta^{(1)\prime\prime}_{\vec k} + \mathcal{H}(1+g)\,\delta^{(1)\prime}_{\vec k} + \left[\mathcal{H}^2 g + (\mathcal{H}g)^{\prime} - \frac{3}{2}\mathcal{H}^2\Omega_{\rm m}\right]\delta^{(1)}_{\vec k} = 0 \, .
\end{equation}

This equation governs the linear growth of matter perturbations in Fourier space and explicitly shows how the dark sector interaction modifies both the effective friction term and the gravitational source responsible for structure formation. In the non-interacting limit, i.e., $g=0$, Eq.~\eqref{eq:ddelta_ids_linear_k} reduces to the standard linear growth equation of the $\Lambda$CDM model, which sets the reference case throughout this work. Assuming separability between time and space, the growing-mode solution of the linear density contrast can be written as,
\begin{equation} \label{eq:deltak_1}
\delta^{(1)}(\vec x,\tau)=D_+(\tau)\,\delta_0(\vec x)\,,
\end{equation}
where $D_+(\tau)$ is the (growing) linear growth factor and $\delta_0(\vec x)$ encodes the initial conditions. This ansatz is well motivated, since the linear evolution equation for $\delta^{(1)}$ in IDS models has exactly the same mathematical structure as in the $\Lambda$CDM case. More precisely, it is a homogeneous second-order differential equation in time, whose coefficients depend only on the background quantities. The presence of the interaction modifies only the time dependence of these coefficients, but does not alter the overall form of the equation. In this sense, all effects of the dark sector interaction are fully captured by the modified time evolution of the growth factor $D_+$. 

A key quantity characterizing the growth of cosmic structures is the linear growth rate,
\begin{equation}
f \equiv \frac{d\ln D_+}{d\ln a} = \frac{1}{\mathcal{H}D_+}\frac{dD_+}{d\tau} \,,
\end{equation}
which measures the rate at which matter density perturbations grow as the Universe expands and therefore provides a direct link between the clustering of matter, the background expansion history, and the underlying theory of gravity. The growth rate plays a central role in LSS observations, as it is directly related to the amplitude of peculiar velocities and thus can be used to assess the strength of redshift-space distortions (RSD) in galaxy surveys. Due to the interaction term in the continuity equation (\ref{eq:delta1_linear_k}), the relation between the density contrast and the velocity divergence is modified with respect to the standard case, yielding
\begin{equation}
\theta^{(1)}(\vec x,\tau) = -\mathcal{H}\,(f+g)\,D_+(\tau)\,\delta_0(\vec x)\, .
\end{equation}
This modification implies that, in IDS models, the peculiar velocity field is no longer determined solely by the growth of matter perturbations, but also carries a direct imprint of the energy exchange between CDM and DE. As a consequence, velocity-related observables, such as RSD, velocity power spectra, and momentum-density correlations, become particularly sensitive probes of dark sector interactions. These observables provide a complementary and potentially powerful laboratory to test IDS scenarios, as they respond to the interaction both through the altered growth history and the modified relation between density and velocity fields. It is therefore convenient to introduce the effective growth rate
\begin{equation} \label{eq:fQ_def}
f_Q \equiv f+g \, ,
\end{equation}
which encapsulates the combined effect of gravitational growth and dark sector interaction on the evolution of the velocity field. With this definition, the formal structure of the linear relations between density and velocity perturbations can be written in close analogy with the standard $\Lambda$CDM case, while all the physical effects of the interaction are absorbed into the modified time dependence of $f_Q$. In this sense, the effective growth rate provides a natural extension of the usual growth rate, allowing the familiar $\Lambda$CDM expressions to be formally recovered in the limit $g\to 0$, while retaining a clear physical interpretation in interacting dark sector models. In Fourier space, the corresponding growing-mode solutions take the separable form
\begin{equation} \label{eq:sol1}
\delta^{(1)}_{\vec k} = D_+(\tau)\,\delta_0(\vec k)\,, \qquad
\theta^{(1)}_{\vec k} = -\mathcal{H}f_Q\,D_+(\tau)\,\delta_0(\vec k)\,,
\end{equation}
which constitute the linear building blocks for the higher-order perturbative expansion developed in the following sections. 

These expressions highlight that, while the mathematical structure of the perturbative framework closely mirrors that of $\Lambda$CDM, the dynamics of both density and velocity perturbations is nontrivially altered by the dark sector interaction through the modified growth factor and effective growth rate. 

\subsection{Second-order perturbations}
\label{ssec:second}

Extending the perturbative expansion to second order allows us to capture the leading nonlinear corrections to the evolution of matter density and velocity perturbations. At this order, the continuity and Euler equations, given by Eqs.~\eqref{eq:continuity_ids} and~\eqref{eq:euler_ids}, acquire quadratic source terms built from products of first-order perturbations, giving rise to mode coupling in Fourier space. These second-order contributions are therefore essential for describing the onset of nonlinear structure formation and constitute the lowest-order corrections entering the matter power spectrum at one-loop level. Retaining only second order terms, the continuity and Euler equations can be written schematically as,
\begin{eqnarray}\label{eq:EDO2_ordem_IDS}
   \delta_{\rm m}^{(2)\prime}+\theta_{\rm m}^{(2)}+\mathcal{H}g\,\delta_{\rm m}^{(2)}&=&-u_{\rm m}^{i(1)}\delta^{(1)}_{\mathrm{m},i}-\delta_{\rm m}^{(1)}\theta_{\rm m}^{(1)} \\
   \theta_{\rm m}^{(2)\prime}+\mathcal{H}\theta_{\rm m}^{(2)}+\frac{3}{2}\mathcal{H}^2\Omega_{\rm m}\delta_{\rm m}^{(2)}&=&-u_{\mathrm{m},i}^{j(1)}u_{\mathrm{m},j}^{i(1)}-u_{\mathrm{m}}^{i(1)}\theta_{\mathrm{m},i}^{(1)}
\end{eqnarray}
where the source terms arise from products of first-order density and velocity fields, given by Eq.~\eqref{eq:sol1}. In contrast to the linear case, these quadratic terms couple different Fourier modes and lead to convolution integrals in momentum space. Working in Fourier space, the second-order continuity and Euler equations take the explicit form,
\begin{eqnarray}
\delta_k^{(2)\prime}+\theta_k^{(2)}+\mathcal{H}g\delta_k^{(2)}&=&\mathcal{H}f_Q D_+^2 \int\frac{d^3k_1\;d^3k_2}{(2\pi)^3}\delta_D(\vec{k}-\vec{k}_1-\vec{k}_2)\alpha(\vec{k}_1,\vec{k}_2)\, \delta_0(\vec{k}_1)\delta_0(\vec{k}_2)\,, \nonumber \\[-0.9mm]
&&\label{eq:delta2k} \\
\theta_k^{(2)\prime}+\mathcal{H}\theta_k^{(2)}+\frac{3}{2}\mathcal{H}^2\Omega_{\rm m}\,\delta_k^{(2)}&=&-\mathcal{H}^2 f_Q^2 D_+^2\int\frac{d^3k_1\;d^3k_2}{(2\pi)^3}\delta_D(\vec{k}-\vec{k}_1-\vec{k}_2)\beta(\vec{k}_1,\vec{k}_2)\delta_0(\vec{k}_1)\delta_0(\vec{k}_2)\,, \nonumber \\[-0.9mm]
&&\label{eq:theta2k}
\end{eqnarray}
where the mode-coupling functions
\begin{equation}
\alpha(\vec{k}_1,\vec{k}_2)=1+\frac{\vec{k}_1\!\cdot\!\vec{k}_2}{k_1^2}\,,
\qquad
\beta(\vec{k}_1,\vec{k}_2)=
\frac{\vec{k}_1\!\cdot\!\vec{k}_2}{k_1^2}
+\frac{(\vec{k}_1\!\cdot\!\vec{k}_2)^2}{k_1^2 k_2^2}
\end{equation}
In the present treatment, $\alpha$ and $\beta$ are identical to their $\Lambda$CDM counterparts because we neglect perturbations of the interaction term. Motivated by the structure of the linear solutions, the second-order density contrast and velocity divergence can be written in separable form as
\begin{equation} \label{eq:AB}
\delta_k^{(2)} = D_+^2(\tau)\,A_\delta(\vec{k},\tau)\,,
\qquad
\theta_k^{(2)} = -\mathcal{H}f_Q D_+^2(\tau)\,B_\theta(\vec{k},\tau)\,,
\end{equation}
where the functions $A_\delta$ and $B_\theta$ describe the remaining scale and time dependence. 
In standard $\Lambda$CDM, these functions are often approximated as time independent under the EdS assumption. 
After factoring out the dominant time dependence $D_+^2(\tau)$, they characterize the second-order density and velocity perturbations and retain the information about the nonlinear coupling between different Fourier modes. 
This structure becomes explicit when the functions are written in terms of convolution integrals involving the second-order perturbative kernels,
\begin{eqnarray}
\delta_k^{(2)} &=& D_+^2(\tau)\int\frac{d^3k_1\;d^3k_2}{(2\pi)^3}\delta_D(\vec{k}-\vec{k}_1-\vec{k}_2)\,
F_2(\vec{k}_1,\vec{k}_2,\tau)\,
\delta_0(\vec{k}_1)\delta_0(\vec{k}_2)\,, \label{eq:F2_def}\\
\theta_k^{(2)} &=& -\mathcal{H}f_Q D_+^2(\tau)
\int\frac{d^3k_1\;d^3k_2}{(2\pi)^3}\delta_D(\vec{k}-\vec{k}_1-\vec{k}_2)\,
G_2(\vec{k}_1,\vec{k}_2,\tau)\,
\delta_0(\vec{k}_1)\delta_0(\vec{k}_2)\,. \label{eq:G2_def}
\end{eqnarray}
In this representation, the kernels $F_2$ and $G_2$ fully encode the scale and configuration dependence of the second-order mode coupling, while the functions $A_\delta$ and $B_\theta$ correspond to the convolution of these kernels with the linear density field. 

Substituting Eqs.~\eqref{eq:F2_def} and~\eqref{eq:G2_def} into Eqs.~\eqref{eq:delta2k} and~\eqref{eq:theta2k} yields a closed system of coupled differential equations governing the evolution of the second-order kernels,
\begin{eqnarray}
\frac{F_2'}{\mathcal{H}f_Q}+\left(\frac{2f+g}{f_Q}\right)F_2-G_2 &=& \alpha\,, \label{eq:F2_ids}\\
-\frac{G_2'}{\mathcal{H}f_Q}
-\left(\frac{3}{2}\frac{\Omega_{\rm m}}{f_Q^2}+\frac{f}{f_Q}\right)G_2
+\frac{3}{2}\frac{\Omega_{\rm m}}{f_Q^2}F_2 &=& -\beta\,. \label{eq:G2_ids}
\end{eqnarray}

Equations~\eqref{eq:F2_ids} and~\eqref{eq:G2_ids} already provide clear intuition that, in IDS context, the second-order kernels $F_2$ and $G_2$ are generically time dependent. To better understand the origin of this behavior, it is instructive to contrast these equations with their $\Lambda$CDM counterparts. In the $\Lambda$CDM limit, where $g=0$ and consequently $f_Q=f$, the coefficient multiplying $F_2$ in Eq.~\eqref{eq:F2_ids} reduces to a constant. In IDS models, however, the same coefficient is $(2f+g)/f_Q$, which is in general time dependent due to the explicit contribution of the interaction function $g$ and its nontrivial interplay with the growth rate $f$. The situation is similar, though slightly more subtle, for the second equation. In the $\Lambda$CDM case, the coefficients in Eq.~\eqref{eq:G2_ids} depend on the combination $\Omega_{\rm m}/f^2$, which is strictly equal to unity in the EdS limit and remains close to a constant in realistic $\Lambda$CDM cosmologies\footnote{In $\Lambda$CDM, the growth rate is well approximated by $f\simeq\Omega_{\rm m}^{0.55}$ over a wide redshift range, implying that $\Omega_{\rm m}/f^2$ varies slowly with time and stays close to unity.}. This property justifies the widespread use of the EdS approximation in standard perturbation theory, even though $\Lambda$CDM does not exactly correspond to an EdS universe. In IDS models, by contrast, the combination $\Omega_{\rm m}/f^2$ is replaced by $\Omega_{\rm m}/f_Q^2$, which is no longer expected to be approximately constant in general. Moreover, the additional term $f/f_Q$ introduces further explicit time dependence in the coefficients of Eq.~\eqref{eq:G2_ids}, reflecting the fact that the growth of density perturbations and the evolution of the velocity field are directly influenced by the dark sector interaction.

This analysis illustrates that, in the $\Lambda$CDM model (particularly in the EdS limit) all coefficients appearing in the kernel evolution equations are constant and, since the mode-coupling functions $\alpha$ and $\beta$ depend only on scale, this is fully consistent with time-independent kernel solutions. In this case, the time derivatives of the kernels vanish in Eqs.~\eqref{eq:F2_ids} and~\eqref{eq:G2_ids}, leading to the usual purely algebraic solutions for the second-order kernels $F_2$ and $G_2$. The situation is fundamentally different in IDS models. In the presence of dark sector interactions, the coefficients of the kernel equations become generically time dependent, while the mode-coupling functions remain purely scale dependent. This mismatch necessarily induces a nontrivial time dependence in the second-order kernels. Retaining this time dependence is therefore essential for an accurate perturbative description of nonlinear structure formation in IDS cosmologies.

\begin{figure}[h]
    \centering
    \includegraphics[width=0.9\linewidth]{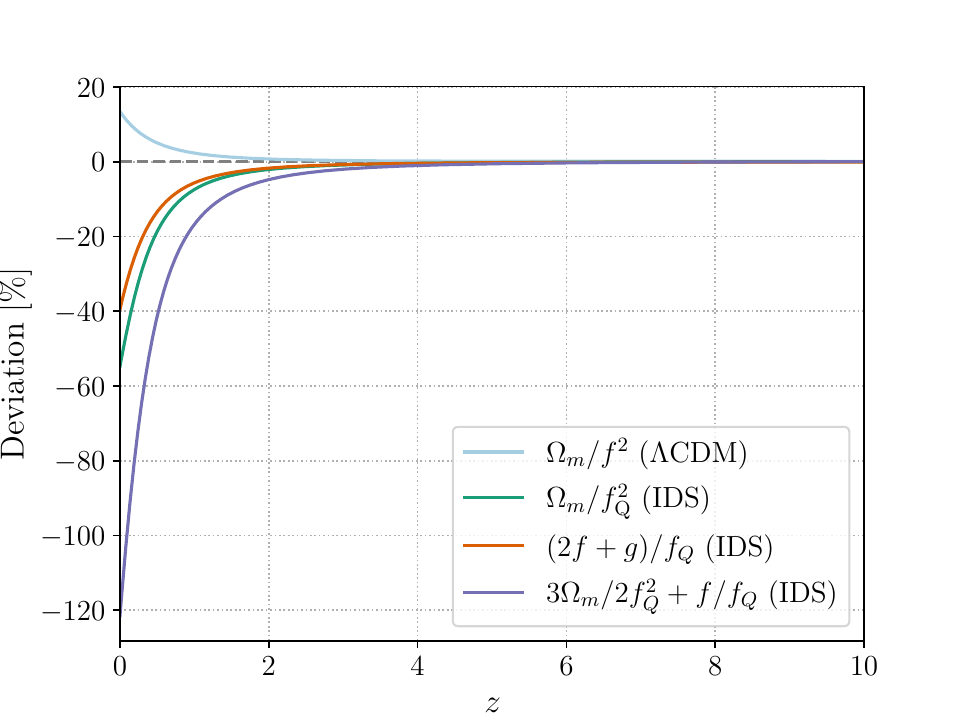}
    \caption{Redshift evolution of the coefficients appearing in the second-order kernel equations, shown relative to their EdS limits. In the $\Lambda$CDM case, the ratio $\Omega_{\rm m}/f^2$ deviates from unity by at most $\sim14\%$. For the IDS model with $\gamma=0.05$, the combinations $\Omega_{\rm m}/f_Q^2$, $(2f+g)/f_Q$, and $(3/2)\,\Omega_{\rm m}/f_Q^2+f/f_Q$ exhibit significantly larger departures from their EdS values, with deviations of at least $\sim40\%$.}
    \label{fig:eds_deviation}
\end{figure}

A more quantitative illustration of this behavior is provided in Fig.~\ref{fig:eds_deviation}, where we compare the redshift evolution of the coefficients appearing in the kernel equations with their corresponding constant values in the EdS limit. Considering first the $\Lambda$CDM case (blue curve), we show that the combination $\Omega_{\rm m}/f^2$, normalized to its EdS value $\Omega_{\rm m}/f^2=1$, remains close to unity over a wide redshift range, deviating noticeably only at $z\lesssim2$ and reaching a maximum deviation of approximately $14\%$ at $z\approx0$. Although this deviation is not negligible at the level of the kernel coefficients, we will show later that its impact on the matter power spectrum is subpercent, which explains the practical success of the EdS approximation in practical applications.

On the other hand, for the IDS scenario, despite the modest value of the interaction parameter, the coefficients entering Eqs.~\eqref{eq:F2_ids} and~\eqref{eq:G2_ids} exhibit significantly larger departures from their EdS limits compared to the $\Lambda$CDM case, starting again around $z\approx2$. More specifically, the green curve shows that the combination $\Omega_{\rm m}/f_Q^2$ deviates from unity by up to $\sim50\%$ at $z\simeq0$, while the orange curve indicates that $(2f+g)/f_Q$ departs from its EdS value of $2$ by approximately $40\%$. Even more remarkably, the combination $(3/2)\,\Omega_{\rm m}/f_Q^2+f/f_Q$ deviates from its EdS limit of $5/2$ by as much as $\sim120\%$ at low redshift. In all cases, as expected, the largest deviations occur at late times, where the effects of the dark sector interaction are most pronounced. 

It is important to emphasize that the quantitative results presented here are model dependent and must be reassessed for different realizations of IDS scenarios. Nevertheless, this analysis demonstrates that even weak, observationally allowed dark sector interactions can induce a substantial time dependence in the second-order kernel coefficients. Whether these enhanced deviations translate into sizable corrections to observable quantities, such as the matter power spectrum, will be addressed explicitly in the following sections.

Finally, since we have not used the symmetrized versions of the mode-coupling functions $\alpha$ and $\beta$ under $\vec{k}_1\leftrightarrow\vec{k}_2$, the kernels obtained from
Eqs.~\eqref{eq:F2_ids} and~\eqref{eq:G2_ids} must be symmetrized. We therefore define
\begin{eqnarray}
F_2^{(s)}(\vec{k}_1,\vec{k}_2,\tau) &=&
\frac{1}{2}\left[F_2(\vec{k}_1,\vec{k}_2,\tau)
+F_2(\vec{k}_2,\vec{k}_1,\tau)\right]\,,\\
G_2^{(s)}(\vec{k}_1,\vec{k}_2,\tau) &=&
\frac{1}{2}\left[G_2(\vec{k}_1,\vec{k}_2,\tau)
+G_2(\vec{k}_2,\vec{k}_1,\tau)\right]\,,
\end{eqnarray}
and, unless stated otherwise, all kernels will be assumed to be symmetrized in the following.

\subsection{Third-order perturbations}
\label{ssec:third}

Proceeding to third order allows us to describe the next level in the nonlinear hierarchy of SPT and, in particular, to obtain the ingredients required for a consistent one-loop computation of the matter power spectrum. At this order, the continuity and Euler equations acquire cubic source terms constructed from products of first- and second-order fields. In Fourier space, these terms couple three wavevectors and can be organized into convolution integrals that define the third-order mode-coupling structure. Retaining only third order terms, the fluid equations can be written schematically as,
\begin{eqnarray} \label{eq:EDO3_ordem_IDS}
\delta_{\rm m}^{(3)\prime}+\theta_{\rm m}^{(3)}+\mathcal{H}g\,\delta_{\rm m}^{(3)}
&=&
-u_{\rm m}^{i(1)}\delta^{(2)}_{{\rm m},i}+u_{\rm m}^{i(2)}\delta^{(1)}_{{\rm m},i}
-\delta_{\rm m}^{(1)}\theta_{\rm m}^{(2)}+\delta_{\rm m}^{(2)}\theta_{\rm m}^{(1)}\,, \\
\theta_{\rm m}^{(3)\prime}+\mathcal{H}\theta_{\rm m}^{(3)}+\frac{3}{2}\mathcal{H}^2\Omega_{\rm m}\delta_{\rm m}^{(3)}
&=&
-u_{{\rm m},i}^{j(1)}u_{{\rm m},j}^{i(2)}+u_{{\rm m},i}^{j(2)}u_{{\rm m},j}^{i(1)}
-u_{\rm m}^{i(1)}\theta^{(2)}_{{\rm m},i}+u_{\rm m}^{i(2)}\theta^{(1)}_{{\rm m},i}\,,
\end{eqnarray}
where the source terms are built from the first- and second-order solutions. Working in Fourier space, the third-order continuity and Euler equations can be cast in the compact convolutional form,
\begin{align}
\delta_{k}^{(3)\prime}+\theta_{k}^{(3)}+\mathcal{H}g\,\delta_{k}^{(3)}
&=\mathcal{H}f_Q D_+^3
\!\!\int\!\!\frac{d^3k_1\,d^3k_2\,d^3k_3}{(2\pi)^6}\,
\delta_D(\vec{k}-\vec{k}_1-\vec{k}_2-\vec{k}_3)\,
\nonumber\\
&\hspace{2.2em}\times
\phi(\vec{k}_1,\vec{k}_2,\vec{k}_3,\tau)\,
\delta_0(\vec{k}_1)\delta_0(\vec{k}_2)\delta_0(\vec{k}_3)\,,
\label{eq:delta3k_compact}\\[0.3em]
\theta_{k}^{(3)\prime}+\mathcal{H}\theta_{k}^{(3)}+\frac{3}{2}\mathcal{H}^2\Omega_{\rm m}\,\delta_{k}^{(3)}
&=-\mathcal{H}^2 f_Q^2 D_+^3
\!\!\int\!\!\frac{d^3k_1\,d^3k_2\,d^3k_3}{(2\pi)^6}\,
\delta_D(\vec{k}-\vec{k}_1-\vec{k}_2-\vec{k}_3)\,
\nonumber\\
&\hspace{2.2em}\times
\psi(\vec{k}_1,\vec{k}_2,\vec{k}_3,\tau)\,
\delta_0(\vec{k}_1)\delta_0(\vec{k}_2)\delta_0(\vec{k}_3)\,.
\label{eq:theta3k_compact}
\end{align}
where $\phi$ and $\psi$ play the role of effective three-mode couplings at third order. They are entirely determined by the standard kinematic factors appearing in the fluid equations together with the second-order kernels $F_2$ and $G_2$. In particular, since $F_2$ and $G_2$ are time dependent in IDS models, the functions $\phi$ and $\psi$ inherit an explicit time dependence. This behavior contrasts with the second-order case, where the mode-coupling functions $\alpha$ and $\beta$ depend only on the wavevectors when perturbations of the interaction are neglected. It also represents a departure from the usual $\Lambda$CDM treatment, where, since $F_2$ and $G_2$ depend only on scale under the EdS approximation, the corresponding third-order couplings are likewise purely scale dependent. A convenient representation is
\begin{eqnarray}
\phi(\vec{k}_1,\vec{k}_2,\vec{k}_3,\tau)
&=&
\alpha(\vec{k}_1,\vec{k}_{23})\,F_2(\vec{k}_2,\vec{k}_3,\tau)
+\alpha(\vec{k}_{12},\vec{k}_3)\,G_2(\vec{k}_1,\vec{k}_2,\tau)\,,
\label{eq:phi_def_compact}\\
\psi(\vec{k}_1,\vec{k}_2,\vec{k}_3,\tau)
&=&
\beta(\vec{k}_1,\vec{k}_{23})\,G_2(\vec{k}_2,\vec{k}_3,\tau)
+\beta(\vec{k}_{12},\vec{k}_3)\,G_2(\vec{k}_1,\vec{k}_2,\tau)\,,
\label{eq:psi_def_compact}
\end{eqnarray}
with $\vec{k}_{ij}\equiv\vec{k}_i+\vec{k}_j$.

Once again, motivated by the structure of the lower-order solutions, we write the third-order density contrast and velocity divergence in terms of third-order kernels,
\begin{eqnarray}
\delta_{k}^{(3)} &=&
D_+^3(\tau)\!\!\int\!\!\frac{d^3k_1\,d^3k_2\,d^3k_3}{(2\pi)^6}\,
\delta_D(\vec{k}-\vec{k}_1-\vec{k}_2-\vec{k}_3)\,
F_3(\vec{k}_1,\vec{k}_2,\vec{k}_3,\tau)\,
\delta_0(\vec{k}_1)\delta_0(\vec{k}_2)\delta_0(\vec{k}_3)\,,
\label{eq:F3_def}\\
\theta_{k}^{(3)} &=&
-\mathcal{H}f_Q D_+^3(\tau)\!\!\int\!\!\frac{d^3k_1\,d^3k_2\,d^3k_3}{(2\pi)^6}\,
\delta_D(\vec{k}-\vec{k}_1-\vec{k}_2-\vec{k}_3)\,
G_3(\vec{k}_1,\vec{k}_2,\vec{k}_3,\tau)\,
\delta_0(\vec{k}_1)\delta_0(\vec{k}_2)\delta_0(\vec{k}_3)\,. \nonumber \\[-0.7em]
&&\label{eq:G3_def}
\end{eqnarray}
The expressions in Eqs.~\eqref{eq:F3_def} and~\eqref{eq:G3_def} make explicit that, as in the second-order case, the dominant time dependence of the third-order perturbations is factored out through $D_+^3(\tau)$, while the kernels $F_3$ and $G_3$ are allowed to retain a residual explicit time dependence that captures departures from the EdS approximation. Moreover, the appearance of the effective growth rate $f_Q$, in place of the standard growth rate $f$, reflects once again that the coupling between density and velocity perturbations at third order is directly modified by the dark sector interaction.

Substituting Eqs.~\eqref{eq:F3_def} and~\eqref{eq:G3_def} into Eqs.~\eqref{eq:delta3k_compact} and~\eqref{eq:theta3k_compact} yields a closed system of coupled differential equations for the third-order kernels,
\begin{eqnarray}
\frac{F_3'}{\mathcal{H}f_Q}+\left(\frac{3f+g}{f_Q}\right)F_3-G_3 &=& \phi\,,
\label{eq:F3_ids}\\
-\frac{G_3'}{\mathcal{H}f_Q}
-\left(\frac{3}{2}\frac{\Omega_{\rm m}}{f_Q^2}+2\frac{f}{f_Q}\right)G_3
+\frac{3}{2}\frac{\Omega_{\rm m}}{f_Q^2}F_3 &=& -\psi\,,
\label{eq:G3_ids}
\end{eqnarray}
where $\phi$ and $\psi$ are given in Eqs.~\eqref{eq:phi_def_compact} and~\eqref{eq:psi_def_compact}. 

In this case, the importance of time dependence is even more pronounced. As can be seen from Eqs.~\eqref{eq:F3_ids} and~\eqref{eq:G3_ids}, coefficients analogous to those appearing at second order are present, however, in addition, the source terms $\phi$ and $\psi$ now also acquire an explicit time dependence through the lower-order kernels $F_2$ and $G_2$. This demonstrates how the interaction-induced time dependence propagates up the perturbative hierarchy, reinforcing the conclusion that a consistent one-loop treatment in interacting dark sector cosmologies requires retaining the full time dependence of the kernels beyond the EdS approximation. Therefore, we do not show explicit curves for the third-order coefficient deviations, as they lead to qualitatively similar behavior to that already illustrated in Fig.~\ref{fig:eds_deviation}.

To ensure invariance under permutations of the wavevectors entering the convolution integrals, we work with fully symmetrized third-order kernels, defined as
\begin{eqnarray}
F_3^{(s)}(\vec{k}_1,\vec{k}_2,\vec{k}_3,\tau) &=&
\frac{1}{3!}\sum_{\pi\in S_3}
F_3(\vec{k}_{\pi(1)},\vec{k}_{\pi(2)},\vec{k}_{\pi(3)},\tau)\,,\\
G_3^{(s)}(\vec{k}_1,\vec{k}_2,\vec{k}_3,\tau) &=&
\frac{1}{3!}\sum_{\pi\in S_3}
G_3(\vec{k}_{\pi(1)},\vec{k}_{\pi(2)},\vec{k}_{\pi(3)},\tau)\,,
\end{eqnarray}
where $S_3$ denotes the permutation group of three elements. Unless explicitly stated otherwise, all third-order kernels will be assumed to be symmetrized in the following.

\subsection{General case: $n$-order perturbations}
\label{ssec:generaln}

Standard Perturbation Theory is most commonly presented directly in its general, all-order formulation, since this compact representation efficiently captures the full nonlinear hierarchy and provides a natural starting point for practical calculations of higher-order observables. In the present work, however, we have chosen to explicitly develop the perturbative expansion up to first, second, and third order. This step-by-step construction serves a pedagogical purpose, since it allows us to clearly identify: ($i$) how the dark sector interactions modify the fluid equations at each perturbative level; ($ii$) how the interaction-induced time dependence enters the kernels; ($iii$) and how these effects propagate through the hierarchy. Having established these features explicitly at low order, we now turn to the general $n$-th order formulation. This general framework will be adopted as the primary working formalism in future papers of this series, where the focus will be on higher-order statistics and phenomenological applications rather than on the explicit derivation of individual perturbative orders.

Collecting only contributions of total perturbative order $n$ in the continuity and Euler equations, given by Eqs.~\eqref{eq:continuity_ids} and~\eqref{eq:euler_ids}, the $n$-th order density contrast and velocity divergence obey the schematic system
\begin{eqnarray}\label{eq:EDO_n_ordem_IDS}
\delta_{\rm m}^{(n)\prime}
+\theta_{\rm m}^{(n)}
+\mathcal{H}g\,\delta_{\rm m}^{(n)}
&=&
S_\delta^{[n]}\!\left[
\left\{
\delta_{\rm m}^{(p)},\theta_{\rm m}^{(q)}
\right\}_{\substack{(p,q<n)\;\land\; (p+q=n)}}
\right],\\
\theta_{\rm m}^{(n)\prime}
+\mathcal{H}\theta_{\rm m}^{(n)}
+\frac{3}{2}\mathcal{H}^2\Omega_{\rm m}\,\delta_{\rm m}^{(n)}
&=&
S_\theta^{[n]}\!\left[
\left\{
\delta_{\rm m}^{(p)},\theta_{\rm m}^{(q)}
\right\}_{\substack{(p,q<n)\;\land\; (p+q=n)}}
\right],
\end{eqnarray}
where $S_\delta^{[n]}$ and $S_\theta^{[n]}$ denote the order-$n$ mode-coupling terms obtained by expanding the nonlinear mode-coupling contributions and retaining only those products of lower-order perturbations whose perturbative orders sum to $n$ (e.g., $\delta^{(m)}\theta^{(n-m)}$).

Analogously to the procedure adopted in the previous sections, it is now convenient to work in Fourier space, where the $n$-th order continuity and Euler equations can be written in a compact convolutional form,
\begin{align}
\delta_{k}^{(n)\prime}+\theta_{k}^{(n)}+\mathcal{H}g\,\delta_{k}^{(n)}
&=\mathcal{H}f_Q D_+^n
\!\!\int\!\!\frac{d^3k_1\cdots d^3k_n}{(2\pi)^{3(n-1)}}\,
\delta_D\!\Big(\vec{k}-\sum_{i=1}^n\vec{k}_i\Big)\,
\Phi_n(\{\vec{k}_i\},\tau)\,
\prod_{i=1}^n\delta_0(\vec{k}_i)\,,
\label{eq:deltank_compact}\\[0.3em]
\theta_{k}^{(n)\prime}+\mathcal{H}\theta_{k}^{(n)}+\frac{3}{2}\mathcal{H}^2\Omega_{\rm m}\,\delta_{k}^{(n)}
&=-\mathcal{H}^2 f_Q^2 D_+^n
\!\!\int\!\!\frac{d^3k_1\cdots d^3k_n}{(2\pi)^{3(n-1)}}\,
\delta_D\!\Big(\vec{k}-\sum_{i=1}^n\vec{k}_i\Big)\,
\Psi_n(\{\vec{k}_i\},\tau)\,
\prod_{i=1}^n\delta_0(\vec{k}_i)\,,
\label{eq:thetank_compact}
\end{align}
where $\Phi_n$ and $\Psi_n$ are effective $n$-mode couplings generated by the quadratic nonlinearities of the fluid equations. The couplings $\Phi_n$ and $\Psi_n$ are built recursively from products of lower-order kernels and therefore inherit an explicit time dependence in IDS models.

Following the structure adopted in the lower-order solutions, the $n$-th order density contrast and velocity divergence can be written in terms of $n$-th order perturbative kernels,
\begin{eqnarray}
\delta_{k}^{(n)} &=&
D_+^n(\tau)\!\!\int\!\!\frac{d^3k_1\cdots d^3k_n}{(2\pi)^{3(n-1)}}\,
\delta_D\!\Big(\vec{k}-\sum_{i=1}^n\vec{k}_i\Big)\,
F_n(\vec{k}_1,\ldots,\vec{k}_n,\tau)\,
\prod_{i=1}^n\delta_0(\vec{k}_i)\,,
\label{eq:Fn_def}\\
\theta_{k}^{(n)} &=&
-\mathcal{H}f_Q D_+^n(\tau)\!\!\int\!\!\frac{d^3k_1\cdots d^3k_n}{(2\pi)^{3(n-1)}}\,
\delta_D\!\Big(\vec{k}-\sum_{i=1}^n\vec{k}_i\Big)\,
G_n(\vec{k}_1,\ldots,\vec{k}_n,\tau)\,
\prod_{i=1}^n\delta_0(\vec{k}_i)\,.
\nonumber\\[-0.7em]
&&\label{eq:Gn_def}
\end{eqnarray}
Accordingly, Eqs.~\eqref{eq:Fn_def} and~\eqref{eq:Gn_def} can be viewed as isolating the dominant time dependence into the overall factor $D_+^n(\tau)$, while the kernels $F_n$ and $G_n$ encode residual, interaction-induced departures from the EdS approximation. Furthermore, the systematic appearance of the effective growth rate $f_Q$ in place of $f$ reflects the fact that the coupling between density and velocity perturbations is modified by the dark sector interaction at every perturbative order.

To make the recursive structure explicit, it is convenient to introduce the partial sums $\vec{k}_{m\cdots n}\equiv \vec{k}_{\rm m}+\cdots+\vec{k}_n$. Then, the effective couplings $\Phi_n$ and $\Psi_n$ can be written in terms of the standard kinematic factors $\alpha$ and $\beta$ and the lower-order kernels as
\begin{eqnarray}
\Phi_n(\vec{k}_1,\ldots,\vec{k}_n,\tau)
&=&
\sum_{m=1}^{n-1}
\alpha\!\left(\vec{k}_{1\cdots m},\vec{k}_{m+1\cdots n}\right)\,
G_m(\vec{k}_1,\ldots,\vec{k}_m,\tau)\,
F_{n-m}(\vec{k}_{m+1},\ldots,\vec{k}_n,\tau)\,, \nonumber \\[-0.9em]
&&\label{eq:Phi_n_def}\\
\Psi_n(\vec{k}_1,\ldots,\vec{k}_n,\tau)
&=&
\sum_{m=1}^{n-1}
\beta\!\left(\vec{k}_{1\cdots m},\vec{k}_{m+1\cdots n}\right)\,
G_m(\vec{k}_1,\ldots,\vec{k}_m,\tau)\,
G_{n-m}(\vec{k}_{m+1},\ldots,\vec{k}_n,\tau)\,. \nonumber \\[-0.9em]
&&
\label{eq:Psi_n_def}
\end{eqnarray}

Finally, substituting Eqs.~\eqref{eq:Fn_def} and~\eqref{eq:Gn_def} into Eqs.~\eqref{eq:deltank_compact} and~\eqref{eq:thetank_compact} leads to a closed system of coupled differential equations governing the evolution of the $n$-th order kernels,
\begin{eqnarray}
\frac{F_n'}{\mathcal{H}f_Q}+\left(\frac{nf+g}{f_Q}\right)F_n-G_n
&=&
\Phi_n(\vec{k}_1,\ldots,\vec{k}_n,\tau)\,,
\label{eq:Fn_ids}\\
-\frac{G_n'}{\mathcal{H}f_Q}
-\left(\frac{3}{2}\frac{\Omega_{\rm m}}{f_Q^2}+(n-1)\frac{f}{f_Q}\right)G_n
+\frac{3}{2}\frac{\Omega_{\rm m}}{f_Q^2}F_n
&=&
-\Psi_n(\vec{k}_1,\ldots,\vec{k}_n,\tau)\,,
\label{eq:Gn_ids}
\end{eqnarray}
where $\Phi_n$ and $\Psi_n$ are given in Eqs.~\eqref{eq:Phi_n_def} and~\eqref{eq:Psi_n_def}. 

At this point, it is important to state that Eqs.~\eqref{eq:Fn_ids} and~\eqref{eq:Gn_ids} provide the general evolution equations for the $n$-th order perturbative kernels in IDS cosmologies and constitute a natural extension of standard SPT to this class of models. Their structure makes explicit that, within the IDS framework, time dependence plays a central role and cannot be neglected, as is often done in the EdS approximation, without a detailed and model-dependent analysis. The cumulative nature of the time dependence in IDS models is manifest at two levels. First, the coefficients of the homogeneous terms depend explicitly on time through the modified growth rate $f_Q$ and the interaction function $g$, reflecting the direct impact of energy exchange in the dark sector on the linear growth of perturbations. Second, the source terms $\Phi_n$ and $\Psi_n$ inherit additional time dependence through their dependence on lower-order kernels, which are themselves time dependent in IDS cosmologies. This hierarchical propagation of time dependence implies that interaction-induced effects are progressively amplified at higher orders, making the assumption of time-independent kernels inconsistent in general. Retaining the full temporal evolution of the kernels is therefore essential for a consistent perturbative description of nonlinear structure formation in IDS scenarios.

In the limit of vanishing interaction, $g\to 0$, Eqs.~\eqref{eq:Fn_ids} and \eqref{eq:Gn_ids} smoothly reduce to the corresponding $\Lambda$CDM hierarchy with time-dependent kernels. In this case, although the coefficients of the homogeneous terms remain time dependent through the combination $\Omega_{\rm m}/f^2$, their evolution is sufficiently slow that the EdS approximation provides an accurate description over a wide range of redshifts. Under this approximation, the coefficients become effectively constant and the kernels reduce to the familiar time-independent solutions obtained from purely algebraic recursion relations. 

As has been done in the specific cases, it is necessary to ensure that the kernels entering the convolution integrals are invariant under permutations of their wavevector arguments. This requirement follows from the symmetry of the underlying correlation functions with respect to the relabeling of Fourier modes. Since the functions $\Phi_n$ and $\Psi_n$, and consequently the kernels $F_n$ and $G_n$ obtained from Eqs.~\eqref{eq:Fn_ids} and~\eqref{eq:Gn_ids}, are not, in general, symmetric under permutations of $\{\vec{k}_i\}$, an explicit symmetrization must be performed. We therefore define the fully symmetrized kernels
\begin{eqnarray}
F_n^{(s)}(\vec{k}_1,\ldots,\vec{k}_n,\tau)
\equiv
\frac{1}{n!}\sum_{\pi\in S_n}
F_n(\vec{k}_{\pi(1)},\ldots,\vec{k}_{\pi(n)},\tau)\,, \\
G_n^{(s)}(\vec{k}_1,\ldots,\vec{k}_n,\tau)
\equiv
\frac{1}{n!}\sum_{\pi\in S_n}
G_n(\vec{k}_{\pi(1)},\ldots,\vec{k}_{\pi(n)},\tau)\,,
\end{eqnarray}
where $S_n$ denotes the permutation group of $n$ elements. Once again, unless stated otherwise, all higher-order kernels will be assumed to be symmetrized in the following.

\section{Results}
\label{sec:results}

Having established the theoretical framework for SPT in IDS models and derived the evolution equations for the time-dependent perturbative kernels, we now turn to their numerical implementation and physical consequences. In this section, we present the main results of the paper, focusing first on the explicit time evolution of the perturbative kernels and subsequently on their impact on the nonlinear matter power spectrum. As previously mention in Sec.~\ref{ssec:beyondbg}, the six standard $\Lambda$CDM cosmological parameters are fixed throughout to the Planck 2018 best-fit values, while the interaction strength is parametrized by $\gamma=0.05$. All background and linear-order quantities, including the modified growth rate $f_Q$ and the linear matter power spectrum, are computed using a suitably modified version of the \texttt{CLASS} Boltzmann code that consistently incorporates the IDS model at both the background and perturbative levels.

\subsection{Time-dependent kernels}
\label{ssec:kernels}

In this subsection we present our numerical solution of the time-dependent second and third order kernels derived in the previous sections. Our strategy is based on a recursive and iterative grid-based approach to solve the coupled system of ordinary differential equations for the kernels. We begin by solving the second-order system for $F_2$ and $G_2$, given by Eqs.~\eqref{eq:F2_ids} and~\eqref{eq:G2_ids}. The equations are integrated on a three-dimensional grid spanning wavenumber and redshift, with $k \in [10^{-3},10]\,\mathrm{Mpc}^{-1}$ and $z \in [0,50]$. Once the numerical solution is obtained on the grid, the kernels are interpolated to arbitrary configurations as required by the convolution integrals entering higher-order calculations. The resulting $F_2$ and $G_2$ kernels are then used as inputs for the third-order mode-coupling functions, allowing us to solve the coupled system for $F_3$ and $G_3$, given by Eqs.~\eqref{eq:F3_ids} and~\eqref{eq:G3_ids}. Since the third-order source terms explicitly depend on the second-order kernels, this procedure naturally defines a recursive hierarchy. We apply the same grid-based integration and interpolation strategy to obtain the final third-order kernels, ensuring a consistent treatment of the time dependence at each perturbative order.

\begin{figure}[h]
    \centering
    \includegraphics[width=\linewidth,trim=10pt 10pt 10pt 10pt,clip]{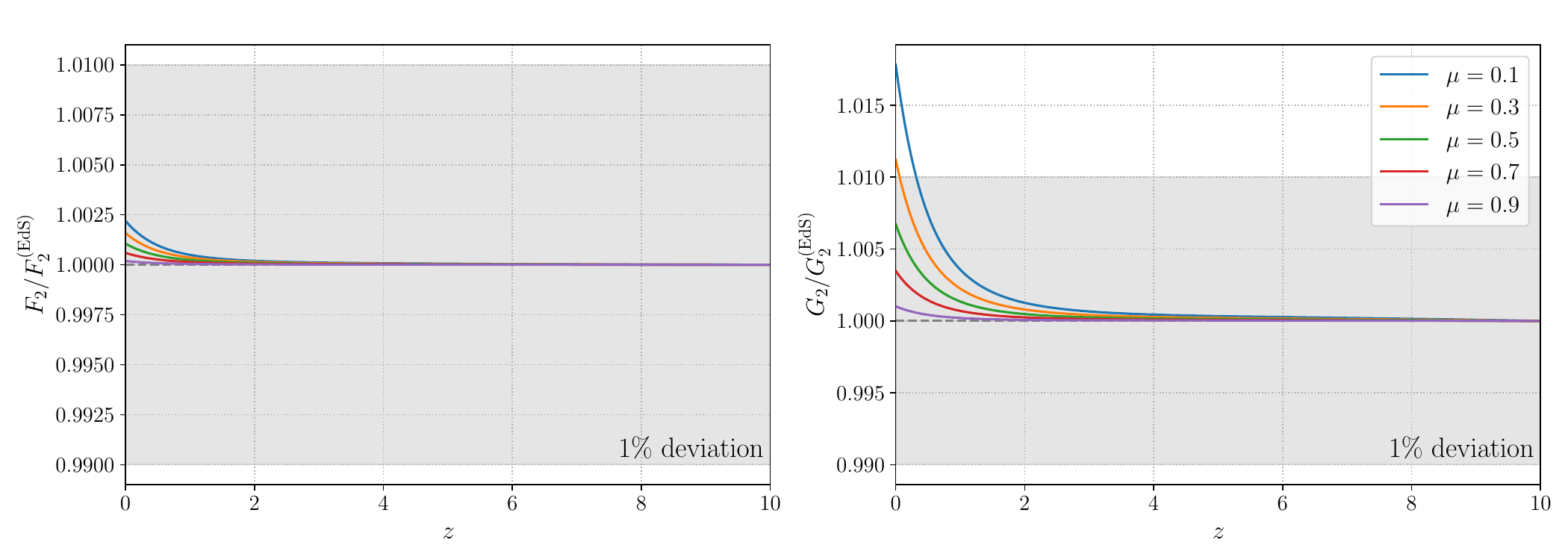}
    \caption{Deviation of the second-order kernels $F_2$ (left) and $G_2$ (right) in $\Lambda$CDM model relative to their EdS counterparts, shown as a function of redshift for the equilateral configuration ($r=1$).}
    \label{fig:F2G2_LCDM_EdS}
\end{figure}

Before turning to IDS models, we first consider the $\Lambda$CDM case. This serves both as a validation of our numerical implementation and as a benchmark against the standard EdS kernels. In Fig.~\ref{fig:F2G2_LCDM_EdS}, we show the ratio of the numerically computed $F_2$ and $G_2$ kernels in $\Lambda$CDM to their EdS counterparts, focusing on the equilateral configuration, $k_1=k_2$ ($r=1$)\footnote{To guide the eye, we highlight in the figure a shaded grey band corresponding to a $1\%$ deviation from the EdS prediction.}. While this choice suffices to illustrate the main features, a more comprehensive exploration of triangle shapes is clearly required. We therefore present a broader set of configurations in Appendix~\ref{ap:kernel}, where we consider the ranges $r = \{0.1,\,1,\,2,\,5,\,10\}$ and $\mu = \{0.1,\,0.3,\,0.5,\,0.7,\,0.9\}$, covering squeezed, equilateral, and elongated triangles over a wide range of relative orientations. In all configurations analyzed, including in Fig.~\ref{fig:F2G2_LCDM_EdS}, the deviations from the EdS kernels remain small across the full redshift interval considered, typically at or below the percent level.

\begin{figure}[h]
    \centering
    \includegraphics[width=\linewidth,trim=10pt 10pt 10pt 10pt,clip]{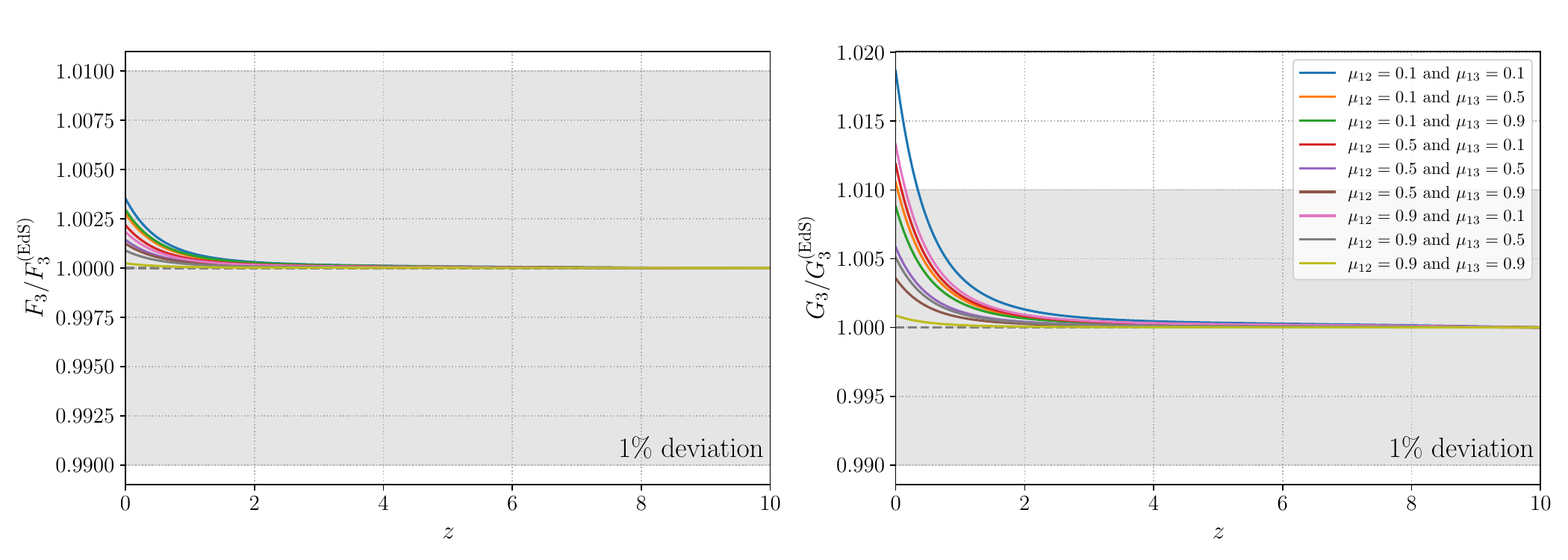}
    \caption{Deviation of the second-order kernels $F_3$ (left) and $G_3$ (right) in $\Lambda$CDM model relative to their EdS counterparts, shown as a function of redshift for the equilateral configuration ($r_2=1$ and $r_3=1$).}
    \label{fig:F3G3_LCDM_EdS}
\end{figure}

We now extend the analysis to the third-order kernels. In Fig.~\ref{fig:F3G3_LCDM_EdS}, we show the ratios of the numerically computed $F_3$ and $G_3$ kernels in the $\Lambda$CDM case to their EdS counterparts, focusing on the equilateral configuration with $k_1=k_2=k_3$ (i.e.\ $r_2=r_3=1$). For this configuration, we probe different relative orientations by varying the angles between the wavevectors, considering $\mu_{12},\,\mu_{13}\in\{0.1,\,0.5,\,0.9\}$. As in the second-order case, the $F_3$ corrections remain sub-percent, while the $G_3$ corrections are of order $\sim 1\%$ over the entire redshift range. In contrast to the second-order analysis, we do not pursue a detailed exploration of different triangle configurations at third order. This choice is motivated by the increased complexity of the vector configurations entering the third-order kernels and by the fact that, when the second-order kernels $F_2$ and $G_2$ are well approximated as time-independent functions, the third-order kernels $F_3$ and $G_3$ can be consistently constructed from $F_2$ and $G_2$, as is commonly done in the EdS limit. Given this recursive structure, and since $F_2$ and $G_2$ have already been analyzed for a variety of configurations in Appendix~\ref{ap:kernel}, we restrict our study of $F_3$ and $G_3$ to a single illustrative configuration and refrain from performing a more exhaustive scan of the parameter space.

These results confirm that, although the $\Lambda$CDM kernels are formally time dependent, the EdS approximation provides an excellent description of their evolution. The result serves as well as a sanity check of our numerical pipeline and validates the accuracy of the recursive grid and interpolation procedure adopted in this work. 

\begin{figure}[h]
    \centering
    \includegraphics[width=\linewidth]{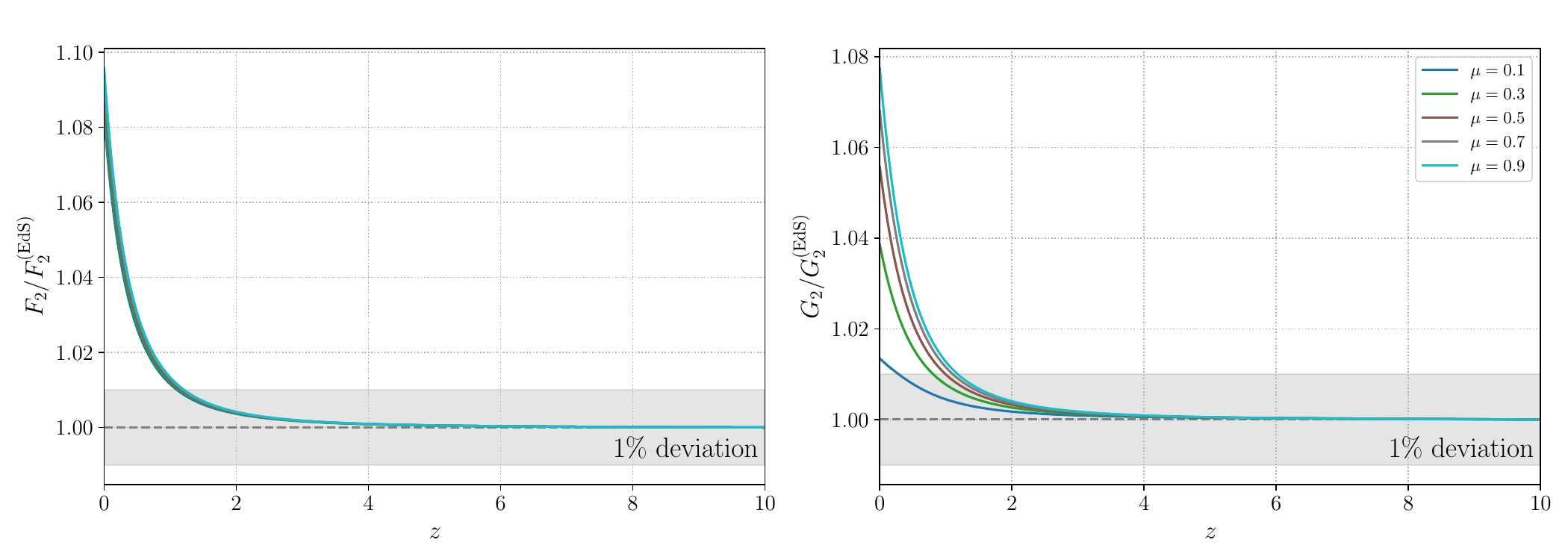}
    \caption{Deviation of the second-order kernels $F_2$ (left) and $G_2$ (right) in IDS models relative to their EdS counterparts, shown as a function of redshift for the equilateral configuration ($r=1$).}
    \label{fig:F2G2_IDS_EdS}
\end{figure}

We now repeat the same analysis for the IDS scenario, using exactly the same set of triangle configurations adopted in the $\Lambda$CDM validation. As in the previous case, Fig.~\ref{fig:F2G2_IDS_EdS} shows the ratios of the numerically computed $F_2$ and $G_2$ kernels in the IDS model to their corresponding EdS counterparts as functions of redshift, focusing on the equilateral configuration. Results for the remaining triangle shapes and orientations are presented in Appendix~\ref{ap:kernel}. In contrast to the $\Lambda$CDM case, the IDS kernels no longer exhibit sub-percent deviations. Instead, depending on the triangle shape and orientation, the departures reach the few-percent level and can approach $\sim 10\%$ at low redshift. 

\begin{figure}[h]
    \centering
    \includegraphics[width=\linewidth]{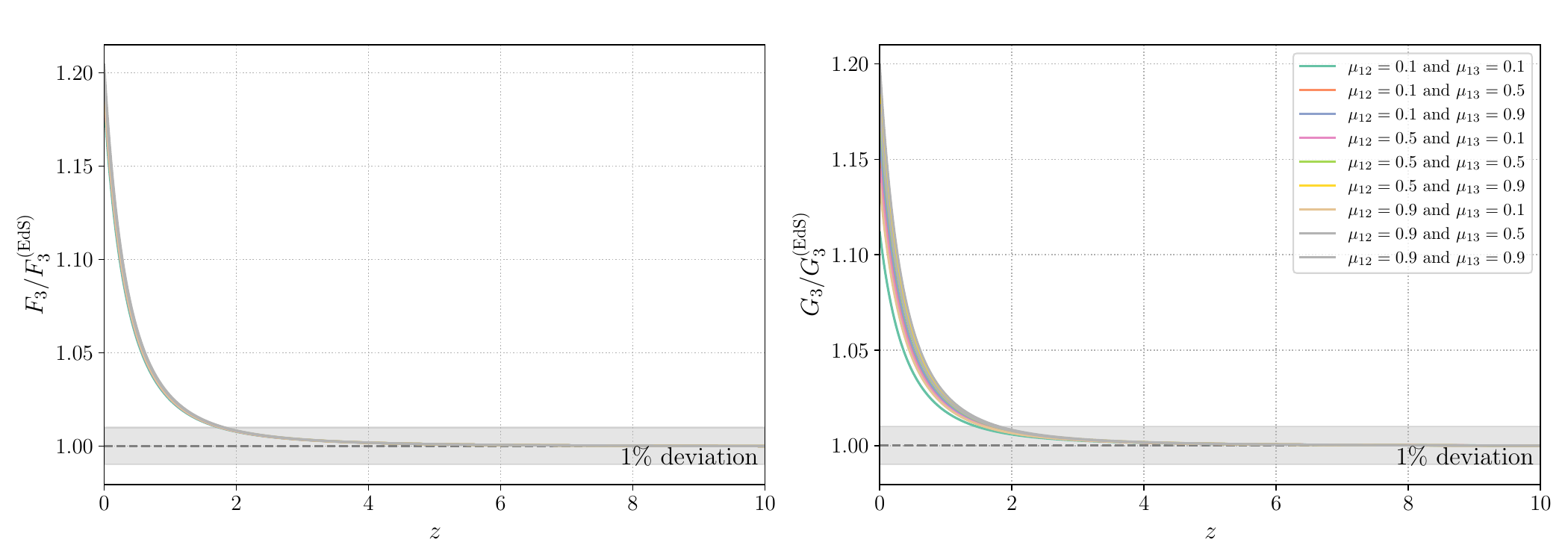}
    \caption{Deviation of the second-order kernels $F_2$ (left) and $G_2$ (right) in IDS models relative to their EdS counterparts, shown as a function of redshift for the equilateral configuration ($r=1$).}
    \label{fig:F3G3_IDS_EdS}
\end{figure}

We now extend this analysis to the third-order kernels in the IDS scenario. In Fig.~\ref{fig:F3G3_IDS_EdS}, we display the ratios of the numerically computed $F_3$ and $G_3$ kernels to their EdS counterparts as functions of redshift, again focusing only on the equilateral configuration with $k_1=k_2=k_3$ ($r_2=r_3=1$) and varying the relative orientations through $\mu_{12}$ and $\mu_{13}$. As before, we do not explicitly present results for additional triangle configurations, since this single representative case is already sufficient to confirm our expectation that time dependence plays a non-negligible role. Compared to the second-order case, the impact of the interaction is even more pronounced at third order. The deviations of $F_3$ and $G_3$ from their EdS limits are generically larger than in $\Lambda$CDM and can reach the $20\%$ level at low redshift. This bigger departure is expected, since at third order the time dependence induced by the interaction enters not only through the kernel evolution coefficients, but also through the source terms, which explicitly depend on the lower-order kernels $F_2$ and $G_2$. As a result, the interaction-driven time dependence propagates cumulatively along the perturbative hierarchy.

These results for the IDS model are fully consistent with the analysis presented in Secs.~\ref{ssec:second} and~\ref{ssec:third}. The presence of a dark sector interaction modifies sufficiently the effective growth rate and introduces explicit time dependence in the coefficients governing the kernel evolution, leading to a breakdown of the EdS approximation of time-independent kernels even for relatively small couplings. At second order, this manifests itself through time-dependent evolution coefficients multiplying $F_2$ and $G_2$, while at third order the effect is further amplified by the explicit time dependence of the source terms, which inherit contributions from the lower-order kernels. The combined behavior of the $F_2$--$G_2$ and $F_3$--$G_3$ systems therefore provides a robust confirmation of the theoretical framework developed in the previous section: the interaction-induced time dependence propagates cumulatively along the perturbative hierarchy and cannot be neglected without loss of accuracy. Physically, this reflects the fact that density and velocity perturbations are no longer governed by a single growth function, but evolve under the continuous exchange of energy between dark matter and dark energy. In that sense, retaining the full time dependence of the perturbative kernels is essential for a consistent one-loop description of nonlinear structure formation in IDS cosmologies.

\subsection{Power spectrum corrections}
\label{ssec:pk}

Having quantified the time dependence of the perturbative kernels, we now turn to its impact on observable quantities by computing the one-loop corrections to the matter power spectrum. The matter power spectrum provides a direct link between the underlying theory of structure formation and LSS observations, and is therefore a natural quantity to assess the physical relevance of interaction-induced modifications. In the framework of SPT, the nonlinear matter power spectrum can be expressed as a perturbative expansion around the linear spectrum,
\begin{equation}
P(k,z)=P_{\rm L}(k,z)+P_{\rm 1\text{-}loop}(k,z)+\cdots,
\end{equation}
where $P_{\rm L}(k,z)$ denotes the linear power spectrum and $P_{\rm 1\text{-}loop}(k,z)$ encodes the leading nonlinear corrections arising from mode coupling. At one-loop order, this contribution is given by the sum of the two standard terms,
\begin{equation}
P_{\rm 1\text{-}loop}(k,z)=P_{22}(k,z)+P_{13}(k,z)\,,
\end{equation}
where $P_{22}(k,z)$ arises from the auto-correlation of second-order density perturbations and captures nonlinear mode coupling between pairs of Fourier modes, while $P_{13}(k,z)$ corresponds to the cross-correlation between first- and third-order contributions and encodes the leading correction due to the coupling between linear and cubic perturbations.

In the present IDS framework, the formal structure of the one-loop corrections remains identical to that of the standard $\Lambda$CDM case, in the sense that the nonlinear power spectrum is still built from the $P_{22}$ and $P_{13}$ contributions. The crucial difference, however, lies in the perturbative kernels entering these loop integrals. As discussed in Sec.~\ref{ssec:kernels}, the kernels are no longer well approximated by their time-independent EdS forms, but are instead given by the fully time-dependent solutions of Eqs.~\eqref{eq:F2_ids},~\eqref{eq:G2_ids},\eqref{eq:F3_ids} and~\eqref{eq:G3_ids}. As a result, the loop corrections explicitly encode the effects of the interaction through the modified growth rate $f_Q$, the coupling function $g$, and the induced time evolution of the kernels $F_n$ and $G_n$ at all relevant orders.

Making use of the kernel definitions introduced in Secs.~\ref{ssec:second} and~\ref{ssec:third}, the one-loop contributions can be written in their standard SPT form as
\begin{eqnarray}
P_{22}(k,z) &=& 2 \int\frac{d^3q}{(2\pi)^3}\,
\left[F_2^{(s)}(\vec{q},\vec{k}-\vec{q},z)\right]^2\,
P_{\rm L}(q,z)\,P_{\rm L}(|\vec{k}-\vec{q}|,z),
\label{eq:P22_def}\\[0.3em]
P_{13}(k,z) &=& 6\,P_{\rm L}(k,z)\int\frac{d^3q}{(2\pi)^3}\,
F_3^{(s)}(\vec{k},\vec{q},-\vec{q},z)\,
P_{\rm L}(q,z),
\label{eq:P13_def}
\end{eqnarray}

It is worth recalling at this point that $P_{\rm L}(k,z)$ denotes the linear matter power spectrum, which is computed consistently for each cosmological model using a suitably modified version of the \texttt{CLASS} Boltzmann code, and that $F_2$ and $F_3$ correspond to the symmetrized perturbative kernels. In particular, the $P_{13}$ contribution depends on the special ``collapsed'' configuration $(\vec{q},-\vec{q})$, highlighting that the third-order kernel enters the one-loop power spectrum only through a restricted subset of configurations. By contrast, the $P_{22}$ term probes the full shape dependence of the second-order kernel, as it involves an integral over all triangle configurations satisfying the closure condition $\vec{k}=\vec{q}+(\vec{k}-\vec{q})$.

\begin{figure}[h]
    \centering
    \includegraphics[width=0.495\linewidth]{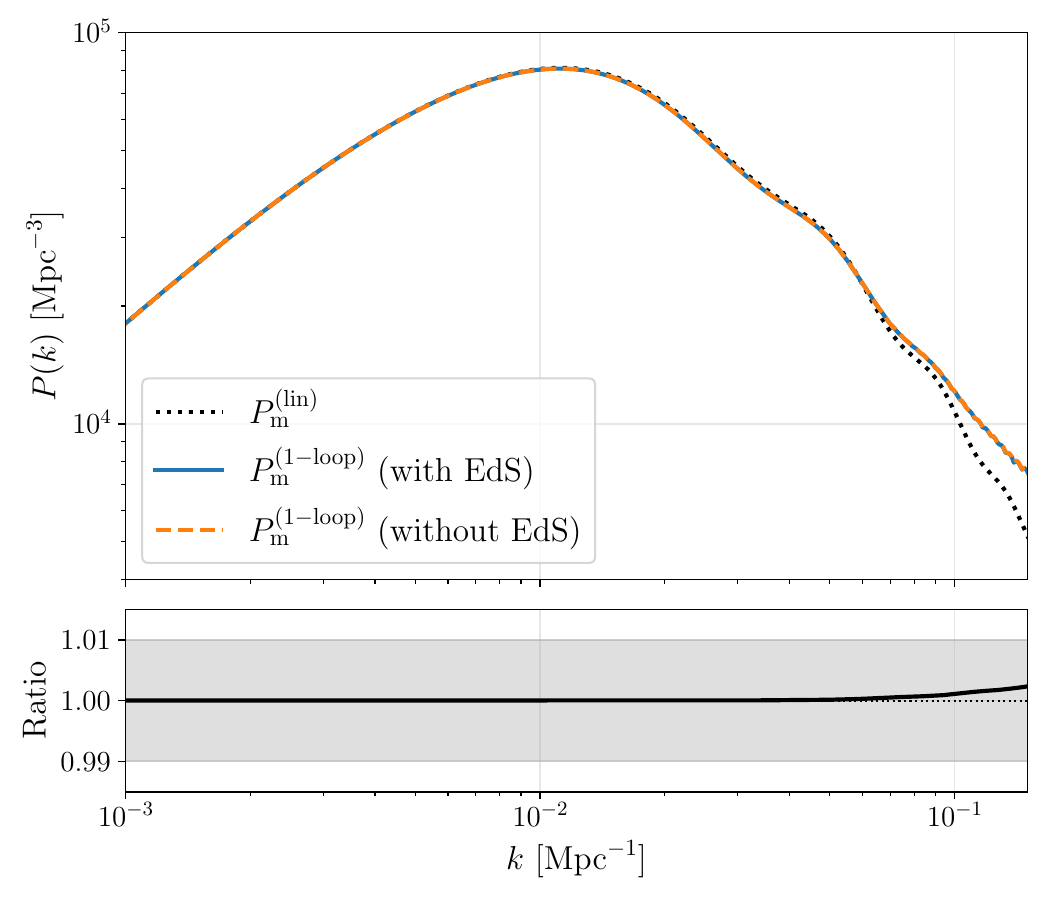}
    \includegraphics[width=0.495\linewidth]{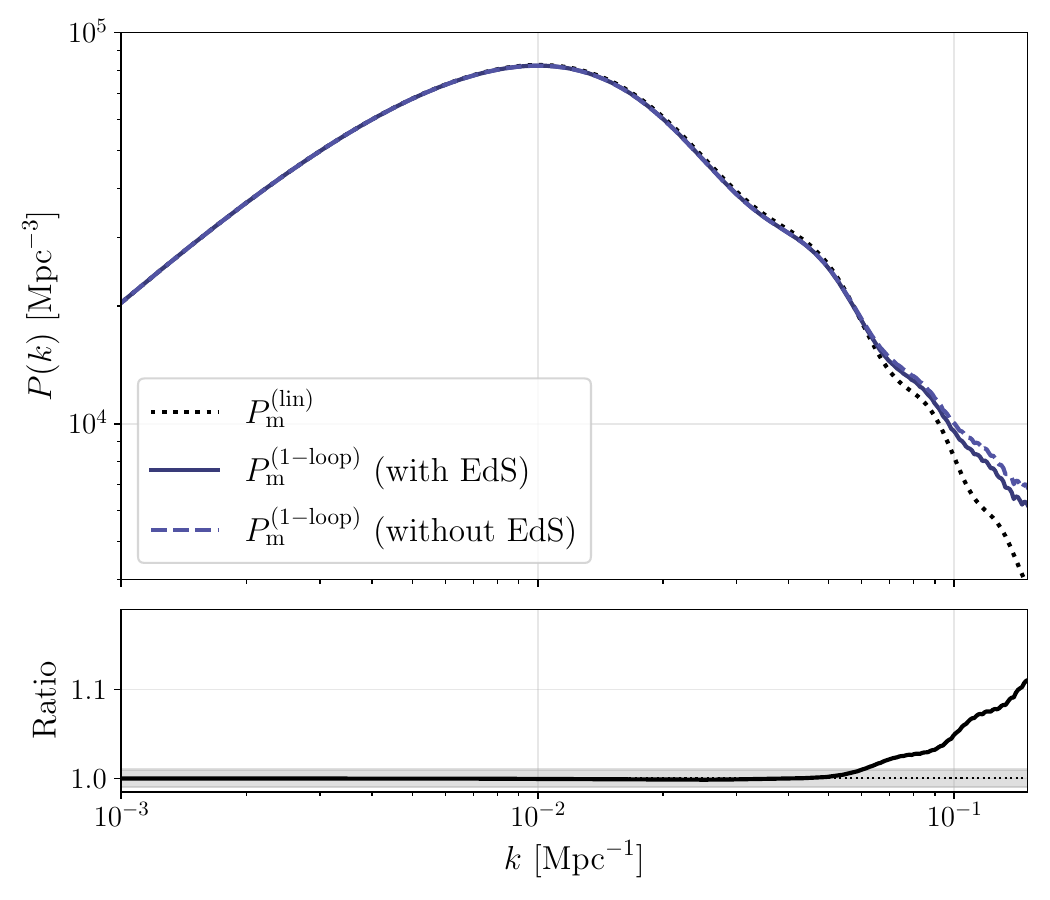}
    \caption{Matter power spectrum and impact of the EdS approximation on the one-loop prediction. \textbf{Left panel:} $\Lambda$CDM model. Right panel: IDS model. In the upper panels, the dotted black curve shows the linear matter power spectrum, while the solid and dashed lines correspond to the one-loop results obtained with and without the EdS kernels, respectively. The lower subpanels display the ratio between the one-loop spectra computed with and without the EdS approximation. The gray shaded band indicates the 1\% deviation region, highlighting the scale dependence and accuracy of the EdS approximation in each model.}
    \label{fig:pk_full}
\end{figure}

As in the analysis of the perturbative kernels, we begin by considering the $\Lambda$CDM case, which serves as a consistency check. In the left panel of Fig.~\ref{fig:pk_full}, we present the nonlinear matter power spectrum for the $\Lambda$CDM model, computed using both the time-dependent kernels and the EdS kernels (top panel), together with the ratio between the two results (bottom panel). As anticipated from the kernel-level analysis, the agreement with the EdS approximation is excellent. The figure shows that employing the time-dependent solution in the $\Lambda$CDM model leads to sub-percent deviations, typically below $0.5\%$ up to $k=0.15\ \mathrm{Mpc}^{-1}$. These findings further confirm that, although the $\Lambda$CDM kernels are formally time dependent, the EdS approximation provides an accurate description of the one-loop power spectrum. This agreement offers a robust sanity check of our numerical pipeline and establishes a reliable baseline against which the IDS results can be meaningfully compared.

Applying the same numerical procedure to the IDS model, we observe a qualitatively different behavior from that found in $\Lambda$CDM. The nonlinear matter power spectrum for this case is shown in the right panel of Fig.~\ref{fig:pk_full}. Here, the impact at one-loop order remains significant. In contrast to the $\Lambda$CDM scenario, the residual difference in the total nonlinear power spectrum relative to the EdS-based result reaches the few-percent level, with deviations of order $\sim 4\%$ at $k=0.1\ \mathrm{Mpc}^{-1}$ and $\sim 10\%$ at $k=0.15\ \mathrm{Mpc}^{-1}$. This level of discrepancy is cosmologically relevant, particularly in the context of current and forthcoming LSS surveys, and provides clear evidence that the EdS approximation is no longer adequate once dark sector interactions are present, even for small coupling strengths.

These results provide clear quantitative evidence that retaining the full time dependence of the perturbative kernels is indispensable for an accurate description of weakly nonlinear structure formation in IDS cosmologies. The interaction modifies not only the overall amplitude but also the detailed scale dependence of the loop corrections, leading to observable effects in the matter power spectrum. 

It is worth noting that, when using the fully time-dependent kernels obtained from the direct numerical integration of Eqs.~\eqref{eq:P22_def} and~\eqref{eq:P13_def} with the numerical time-dependent solutions of Eqs.~\eqref{eq:F2_ids},~\eqref{eq:G2_ids}~\eqref{eq:F3_ids} and~\eqref{eq:G3_ids}, the computation of the full one-loop power spectrum becomes numerically more delicate. In particular, the direct evaluation of the loop integrals can lead to spurious oscillatory features at intermediate and small scales, reflecting numerical instabilities rather than physical effects. This behavior arises because the standard FFT-based techniques, commonly employed in $\Lambda$CDM analyses, rely crucially on the separability and approximate time independence of the EdS kernels, assumptions that no longer hold once the kernels acquire an explicit time dependence. In the present work, we do not attempt to fully resolve this numerical issue, as our primary goal is to demonstrate the physical importance of retaining the time dependence of the perturbative kernels in IDS models. Nevertheless, this limitation highlights the need for a more efficient and robust computational strategy tailored to time-dependent kernels. Addressing this challenge is the main focus of the second paper in this series, where we will develop improved methods for computing the kernels and the corresponding loop integrals with high numerical accuracy. These advances will be essential for a meaningful comparison with observational data and for fully exploiting LSS measurements as probes of IDS cosmologies.

\section{Conclusions}
\label{sec:conclusions}

In this work we have developed a time-dependent formulation of SPT for IDS cosmologies, explicitly avoiding the EdS approximation that is commonly employed in $\Lambda$CDM-based analyses. Starting from the Boltzmann--Poisson system with a defined interaction, we derived the modified continuity and Euler equations governing matter perturbations and constructed the perturbative expansion up to third order, which is the minimum order required for a consistent one-loop computation of the matter power spectrum.

Our analysis demonstrates that, even in the absence of perturbations of the interaction term, a non-vanishing background energy exchange between dark matter and dark energy generically induces a non-trivial time dependence in the SPT kernels. This effect arises from the modified structure of the continuity equation and from the appearance of the effective growth rate $f_Q = f + g$, which alters the coupling between density and velocity perturbations at all orders. As a consequence, the factorization of time and scale dependence underlying the EdS approximation breaks down in IDS cosmologies.

We quantified this breakdown by explicitly analyzing the evolution equations for the second- and third-order kernels. For a representative interaction strength $\gamma = 0.05$, compatible with current observational constraints, we found that the coefficients entering the kernel equations exhibit substantial departures from their EdS limits at low redshift. In particular, while in $\Lambda$CDM the combination $\Omega_{\rm m}/f^2$ deviates from unity by at most $\sim 14\%$ at $z = 0$, in the IDS case the corresponding combinations $\Omega_{\rm m}/f_Q^2$ and $(2f+g)/f_Q$ deviate by $\sim 40$--$50\%$, and the coefficient $(3/2)\Omega_{\rm m}/f_Q^2 + f/f_Q$ can depart from its EdS value by up to $\sim 120\%$ at late times. These results clearly show that even weak interactions can induce large time variations in the perturbative kernels.

We further assessed the phenomenological impact of these effects by computing the one-loop matter power spectrum using the fully time-dependent kernels. We found that nonlinear corrections are systematically enhanced in IDS models relative to $\Lambda$CDM and that the resulting deviations in the matter power spectrum can significantly exceed the percent level on mildly nonlinear scales, even for small interaction strengths. This demonstrates that adopting EdS kernels in IDS cosmologies can lead to biased predictions that are not negligible in the context of current and forthcoming LSS surveys.

The results presented here establish that retaining the full time dependence of the perturbative kernels is not merely a formal improvement, but a necessary ingredient for a consistent and accurate description of nonlinear structure formation in IDS scenarios. The framework developed in this work provides a solid foundation for precision tests of dark sector interactions using nonlinear large-scale structure observables, including higher-order statistics and redshift-space distortions.

This paper constitutes the first step of a broader project. In forthcoming work, we will extend this analysis to a more efficient way to compute the time-dependent kernels, and investigate the implications of IDS physics for higher-order correlators and observational analyses.

\acknowledgments
The authors are grateful to Guilherme Brando and Tiago Castro for useful discussions and valuable comments that contributed to the development of this work.


\appendix

\section{Additional triangle configurations for the kernels}
\label{ap:kernel}

In this Appendix we present an extended set of triangle configurations employed in the analysis of the second-order kernels $F_2$ and $G_2$, complementing the representative cases discussed in Sec.~\ref{ssec:kernels}. For the $F_2$--$G_2$ system, the triangle geometry in Fourier space is parametrized by the ratio of wavevector magnitudes $r \equiv k_2/k_1$ and by the cosine of the enclosed angle, $\mu \equiv \hat{k}_1\!\cdot\!\hat{k}_2$. In order to explore a broad range of configurations, including squeezed, equilateral, moderately elongated, and highly elongated triangles, we consider the following values:
\begin{eqnarray}
    r &=& \{0.1,\,2,\,10\}\,, \label{eq:val_r} \\
    \mu &=& \{0.1,\,0.3,\,0.5,\,0.7,\,0.9\}\,. \label{eq:val_mu}
\end{eqnarray}

\begin{figure}[h]
    \centering
    \includegraphics[width=\linewidth,trim=10pt 10pt 10pt 10pt,clip]{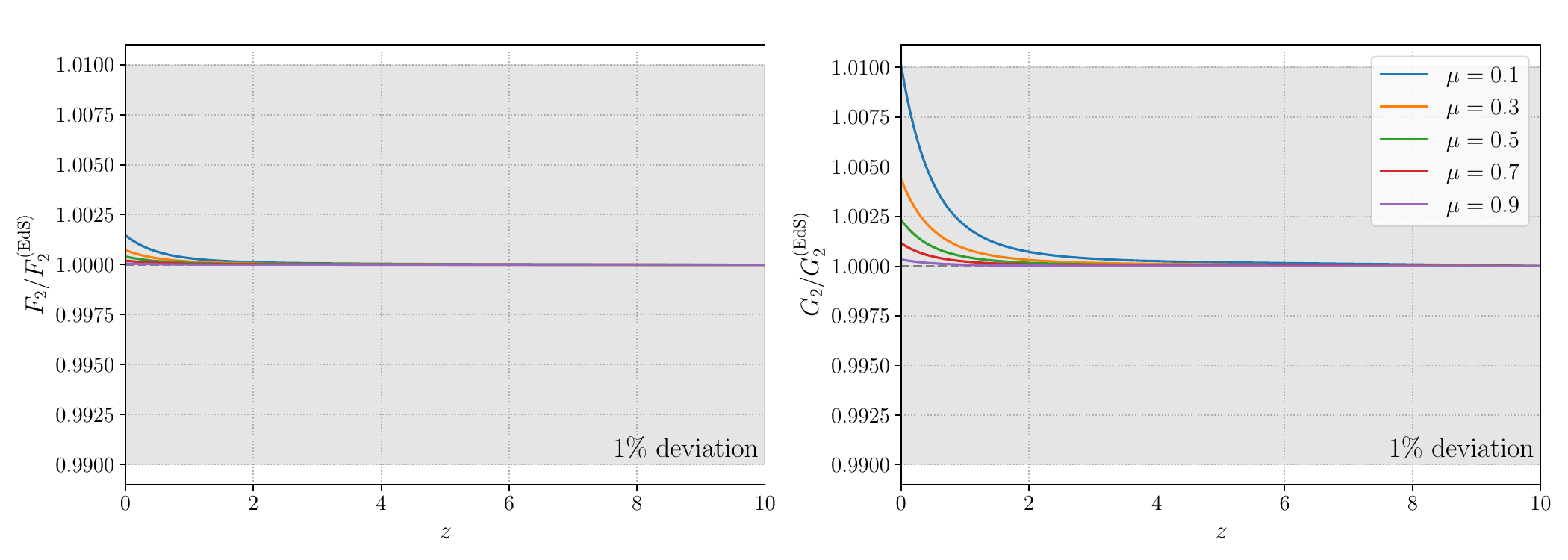}
    \includegraphics[width=\linewidth,trim=10pt 10pt 10pt 10pt,clip]{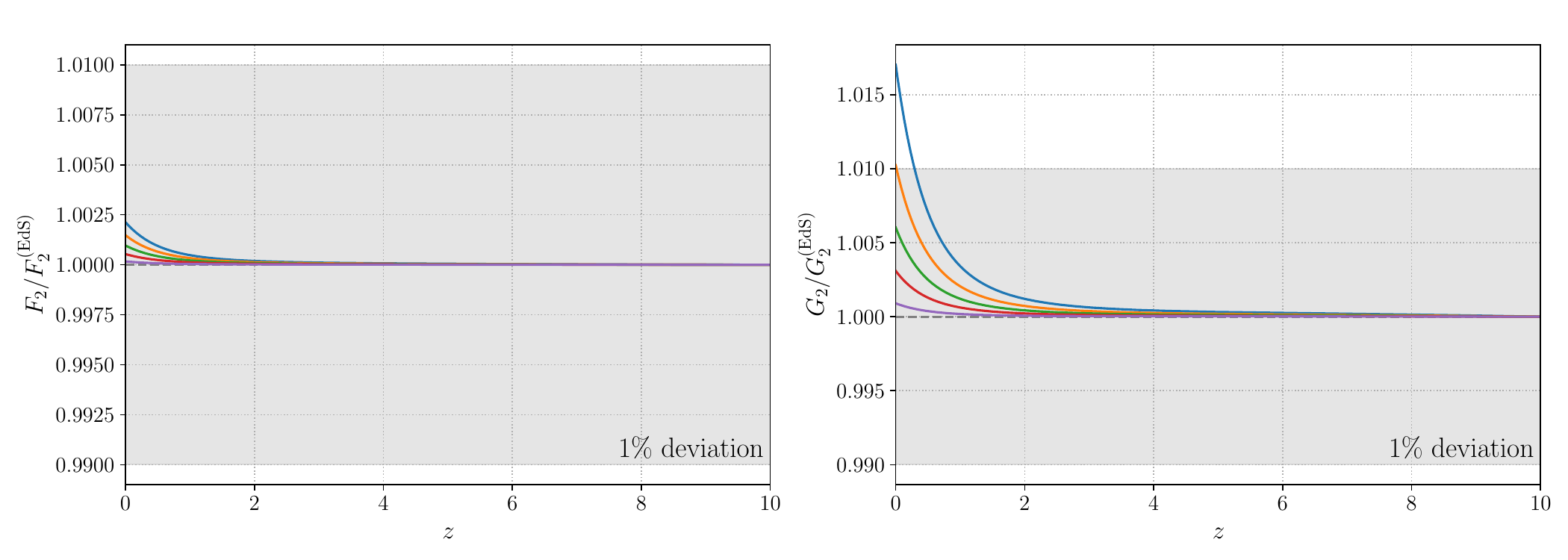}
    \includegraphics[width=\linewidth,trim=10pt 10pt 10pt 10pt,clip]{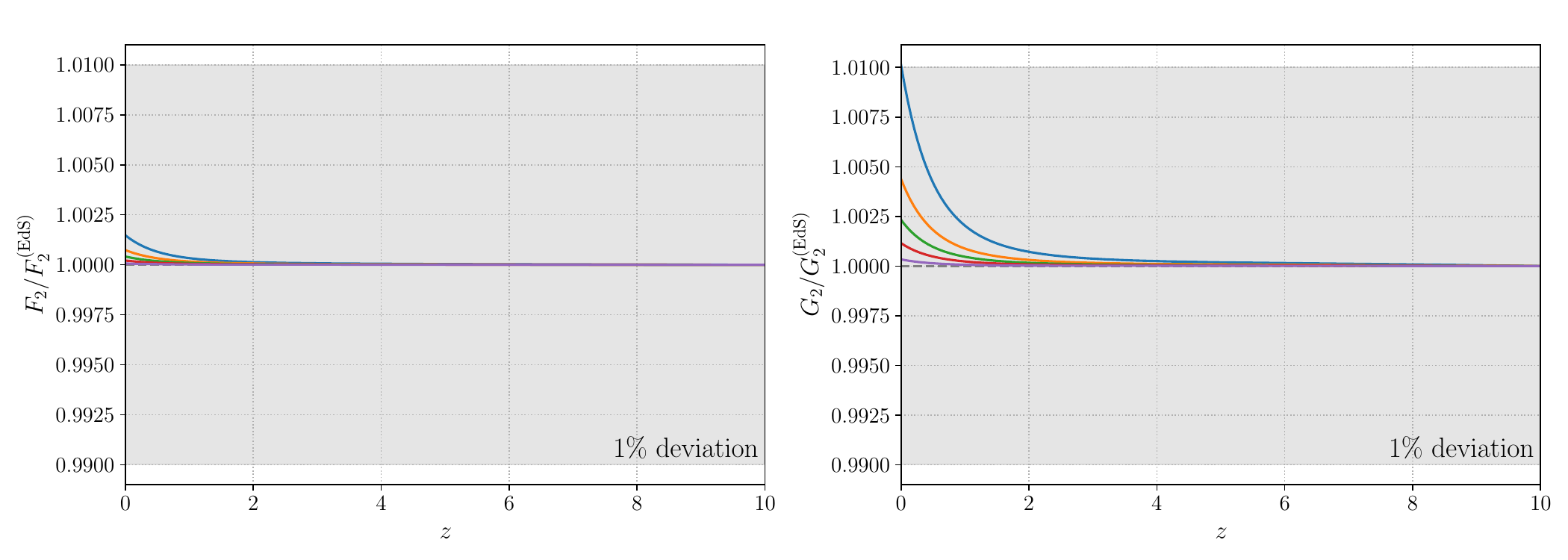}
    \caption{Same as Fig.~\ref{fig:F2G2_LCDM_EdS}, but for additional triangle configurations. 
\textbf{(i) Top row:} $r=0.1$ (squeezed configuration). 
\textbf{(ii) Middle row:} $r=2$ (moderately elongated configuration). 
\textbf{(iii) Bottom row:} $r=10$ (highly elongated configuration).}
    \label{fig:F2G2_LCDM_EdS_vals}
\end{figure}

We begin by presenting the $\Lambda$CDM results. In Fig.~\ref{fig:F2G2_LCDM_EdS_vals}, we show the ratios of the numerically computed second-order kernels to their corresponding EdS limits as functions of redshift for this extended set of triangle configurations. For all triangle shapes and orientations considered, the deviations from the EdS kernels remain small, typically at or below the percent level. This confirms that, although the $\Lambda$CDM kernels are formally time dependent, the EdS approximation provides an excellent description over the full redshift range of interest. This additional figure therefore reinforce the conclusions drawn in Sec.~\ref{ssec:kernels} and demonstrate that they are robust against variations in triangle geometry.

\begin{figure}[h]
    \centering
    \includegraphics[width=\linewidth]{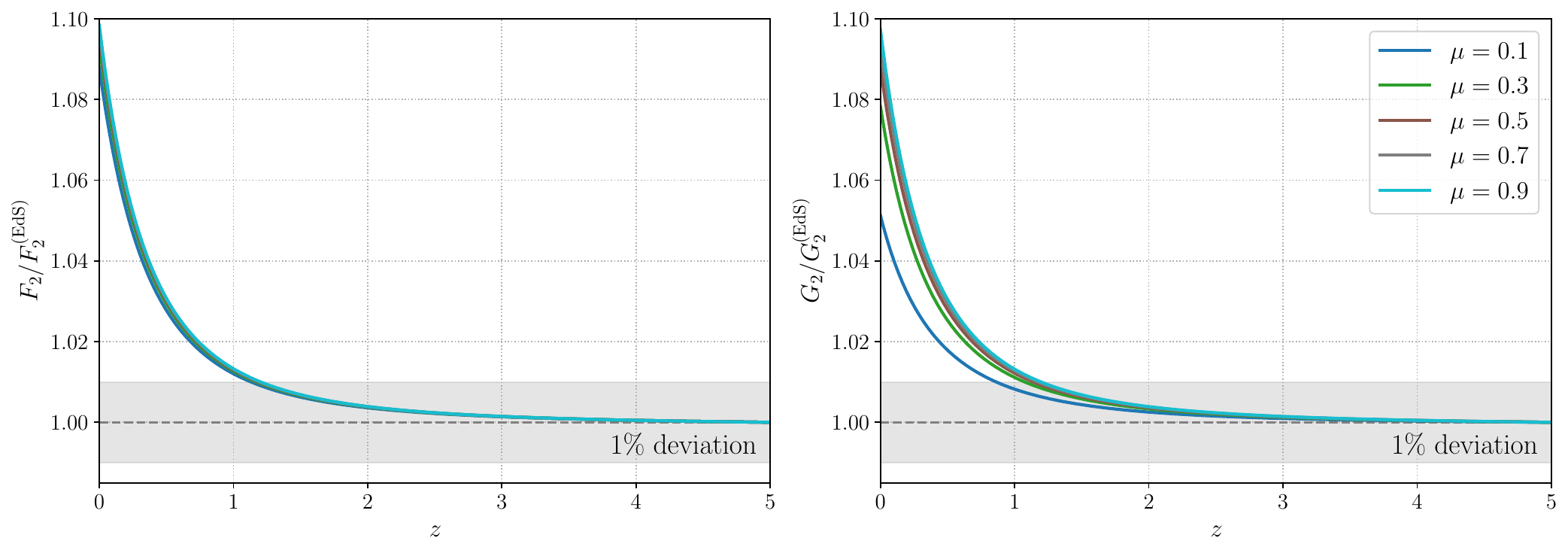}
    \includegraphics[width=\linewidth]{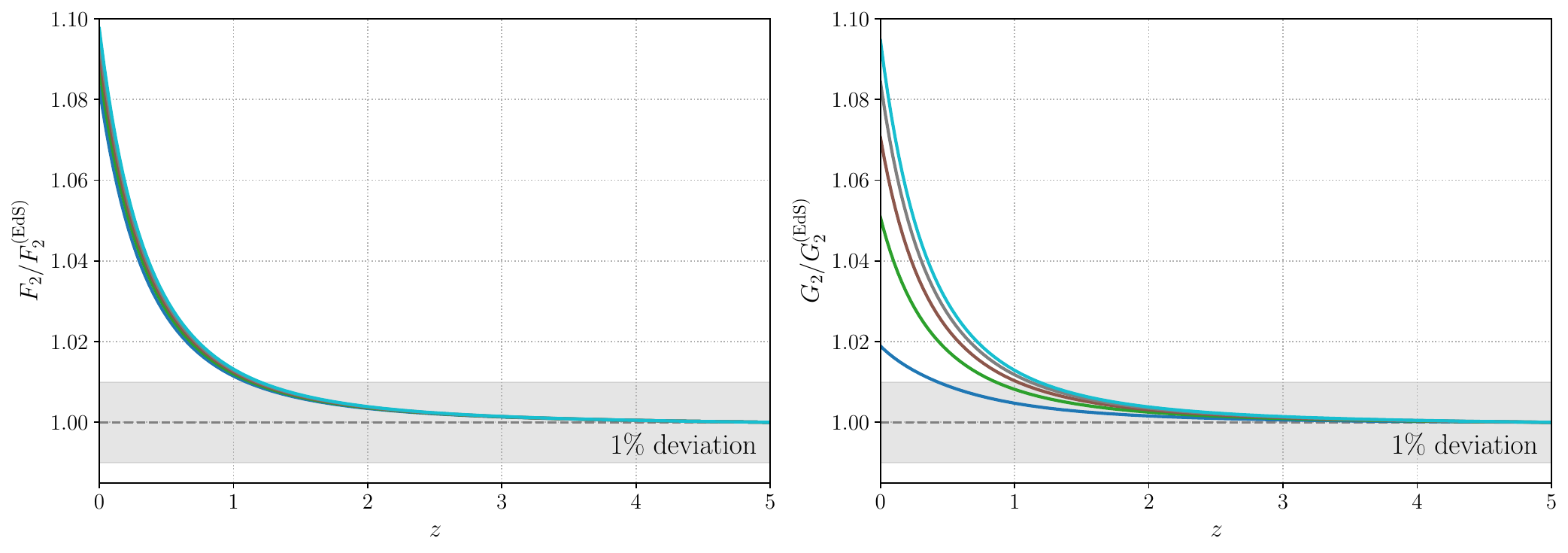}
    \includegraphics[width=\linewidth]{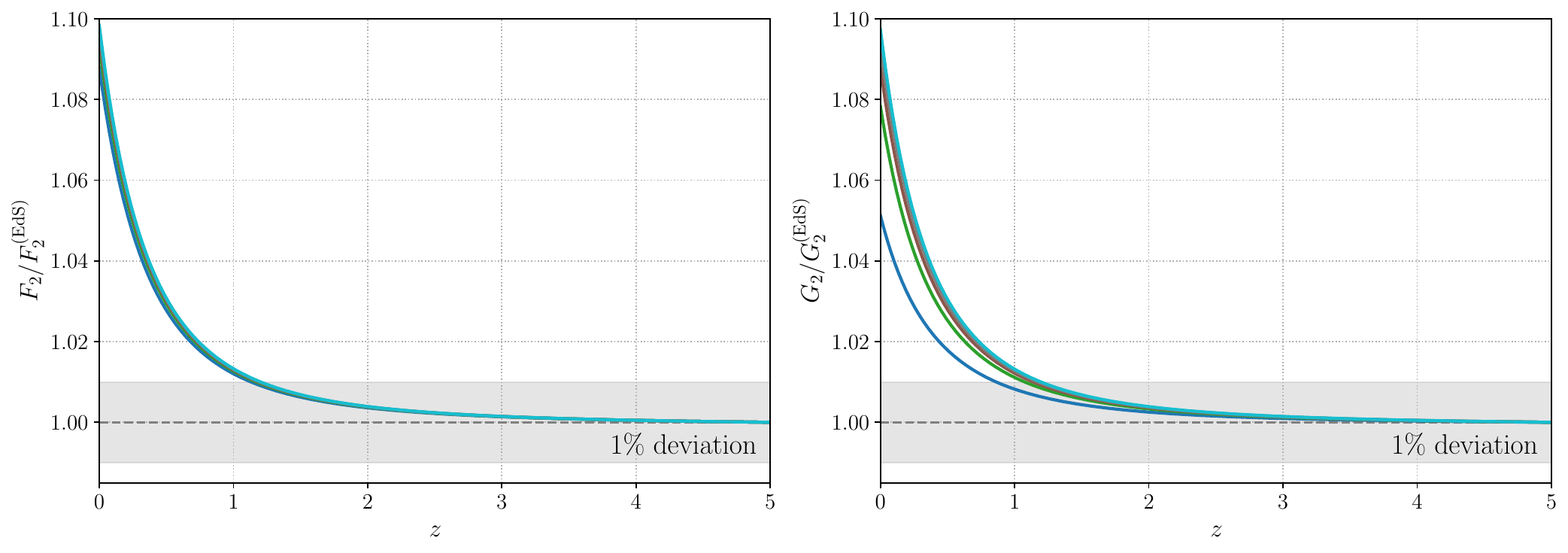}
    \caption{Same as Fig.~\ref{fig:F2G2_IDS_EdS}, but for additional triangle configurations in the IDS scenario. 
    \textbf{(i) Top row:} $r=0.1$ (squeezed configuration). 
    \textbf{(ii) Middle row:} $r=2$ (moderately elongated configuration). 
    \textbf{(iii) Bottom row:} $r=10$ (highly elongated configuration).}
    \label{fig:F2G2_IDS_EdS_vals}
\end{figure}

We now turn to the IDS scenario. In Fig.~\ref{fig:F2G2_IDS_EdS_vals}, we show the ratios of the numerically computed second-order kernels to their corresponding EdS limits as functions of redshift for the same extended set of triangle configurations adopted in the $\Lambda$CDM analysis. In clear contrast with the standard case, the IDS kernels display substantially larger departures from the EdS approximation. The deviations exhibit a pronounced dependence on redshift, reaching the few-percent level for several configurations and approaching $10\%$ at low redshift in almost all cases. Confirming the analysis made in Sec.~\ref{sec:results}, this behavior is a clear manifestation of the interaction-induced modification of the effective growth rate $f_Q$ and the resulting explicit time dependence of the kernel evolution coefficients. 


\bibliographystyle{JHEP}
\bibliography{biblio.bib}

\end{document}